\documentclass[10pt,twocolumn,letterpaper]{article}

\usepackage[pagenumbers]{wacv} 

\usepackage{graphicx}
\usepackage{amsmath}
\usepackage{amssymb}
\usepackage{bm}
\usepackage{dsfont}
\usepackage{cite}

\usepackage{graphicx}
\usepackage{float}
\usepackage{multirow}
\usepackage{booktabs}
\usepackage{array}
\usepackage[export]{adjustbox}
\usepackage{diagbox}

\usepackage{tikz}
\usetikzlibrary{positioning, arrows}

\usepackage{algorithm}
\usepackage{algpseudocode}

\algrenewcommand\algorithmiccomment[1]{\(\triangleright\) #1}

\usepackage[pagebackref,breaklinks,colorlinks]{hyperref}

\usepackage[capitalize]{cleveref}
\crefname{section}{Sec.}{Secs.}
\Crefname{section}{Section}{Sections}
\Crefname{table}{Table}{Tables}
\crefname{table}{Tab.}{Tabs.}

\def\wacvPaperID{397} 
\def\confName{WACV}
\def\confYear{2024}

\newcommand{\berm}{\textit{ProM}}

\newcommand{\s}{\ensuremath{\bm{\theta}}}
\newcommand{\x}{\ensuremath{\bm{x}}}
\newcommand{\m}{\ensuremath{\bm{m}}}
\newcommand{\M}{\ensuremath{\bm{M}}}
\newcommand{\prior}{\ensuremath{p_{\bm{M}}}}
\newcommand{\post}{\ensuremath{p_{\bm{M}|\bm{x}}}}

\newcommand{\eqendp}{\,\text{.}} 
\newcommand{\eqendc}{\,\text{,}} 

\DeclareMathOperator*{\argmin}{\arg\min}
\DeclareMathOperator*{\argmax}{\arg\max}

\begin{document}

\title{Constrained Probabilistic Mask Learning for\\Task-specific Undersampled MRI Reconstruction}

\author{
Tobias Weber\textsuperscript{1, 2}\quad Michael Ingrisch\textsuperscript{1, 2}\quad Bernd Bischl\textsuperscript{1, 2}\quad David Rügamer\textsuperscript{1, 2}\\
\textsuperscript{1} LMU Munich \quad \textsuperscript{2} Munich Center for Machine Learning (MCML) \\
{\tt\small tobias.weber@stat.uni-muenchen.de}
}
\maketitle

\begin{abstract}

Undersampling is a common method in Magnetic Resonance Imaging (MRI) to subsample the number of data points in k-space, reducing acquisition times at the cost of decreased image quality.
A popular approach is to employ undersampling patterns following various strategies, e.g., variable density sampling or radial trajectories.
In this work, we propose a method that directly learns the undersampling masks from data points, thereby also providing task- and domain-specific patterns.
To solve the resulting discrete optimization problem, we propose a general optimization routine called \berm: A fully probabilistic, differentiable, versatile, and model-free framework for mask optimization that enforces acceleration factors through a convex constraint.
Analyzing knee, brain, and cardiac MRI datasets with our method, we discover that different anatomic regions reveal distinct optimal undersampling masks,
demonstrating the benefits of using custom masks, tailored for a downstream task.
For example, \berm\ can create undersampling masks that maximize performance in downstream tasks like segmentation with networks trained on fully-sampled MRIs.
Even with extreme acceleration factors, \berm\ yields reasonable performance while being more versatile than existing methods, paving the way for data-driven all-purpose mask generation.
\footnote{\textbf{Code:} https://github.com/saiboxx/bernoulli-mri}

\end{abstract}
%



\section{Introduction}

Undersampling plays a crucial role in accelerating the acquisition time of magnetic resonance imaging (MRI) by selectively sampling only specific data points in k-space, a representation of spatial frequency information.
Through undersampling, acquisition times in MRI can be shortened, which in turn can be traded into higher spatial resolution, spatial coverage, or shorter scanning sessions.
The latter promise to reduce patient discomfort and anxiety, particularly for individuals who may experience claustrophobia or have difficulty lying still for extended periods \cite{oztek2020practical}.
Likewise, undersampling techniques can be employed to increase temporal resolution in dynamic imaging enabling real-time MRI \cite{uecker2010real, nayak2022real}, allowing for visualization and monitoring of dynamic processes as they occur.
This is particularly relevant in interventional radiology procedures, where real-time guidance is, among others, crucial e.g. for accurate needle placement \cite{kaplan2002real} or catheter navigation \cite{campbell2017real}.
Without optimized reconstruction, however, reducing the number of acquired data points results in decreased image quality and artifacts like infolding, blurring, or distortions, which can hinder accurate diagnosis and interpretation.
To tackle this issue, various techniques have been developed to enhance reconstructions and produce high-quality images from undersampled data, e.g., via compressed sensing \cite{lustig_sparse_2007} or parallel imaging \cite{deshmane2012parallel, heidemann2003brief}.

Rather than focusing on enhancing the image quality for a predetermined undersampling pattern through image reconstruction, this paper addresses the challenge of identifying the optimal sampling pattern or k-space acquisition mask in terms of reconstruction quality or downstream task performance for a given undersampling ratio.
Previously, in the area of deep learning this challenge has been mainly addressed by combined approaches that simultaneously learn a reconstruction network and an undersampling mask \cite{bahadir_learning-based_2019, weiss_joint_2020, xie_puert_2022, wang_b-spline_2022, weiss2019pilot}.
While image enhancement has proven to work well using such neural network approaches, these black box algorithms give little to no control over the reconstruction process and often have practical as well as legal limitations when being used as the basis of a reliable system in safety-critical applications.
Furthermore, the application of large architectures in clinical practice, especially for real-time systems, requires computational resources not affordable by many healthcare institutions, particularly in developing regions where such advanced technology is scarce or cost-prohibitive.  

Most closely related to our work, \cite{razumov_optimal_2022, razumov_optimal_2022-1} propose a direct undersampling mask optimization scheme based on iterative gradient sampling (IGS), which repeatedly determines k-space elements that contribute the most to a loss criterion.
\begin{figure*}[t]
    \centering
    \begin{tikzpicture}[
         image/.style = {text width=\textwidth, 
                         inner sep=0pt, outer sep=0pt},
        node distance = 1mm and 1mm
                                ] 
        \node [] (frame1)
            {\includegraphics[width=0.85\textwidth]{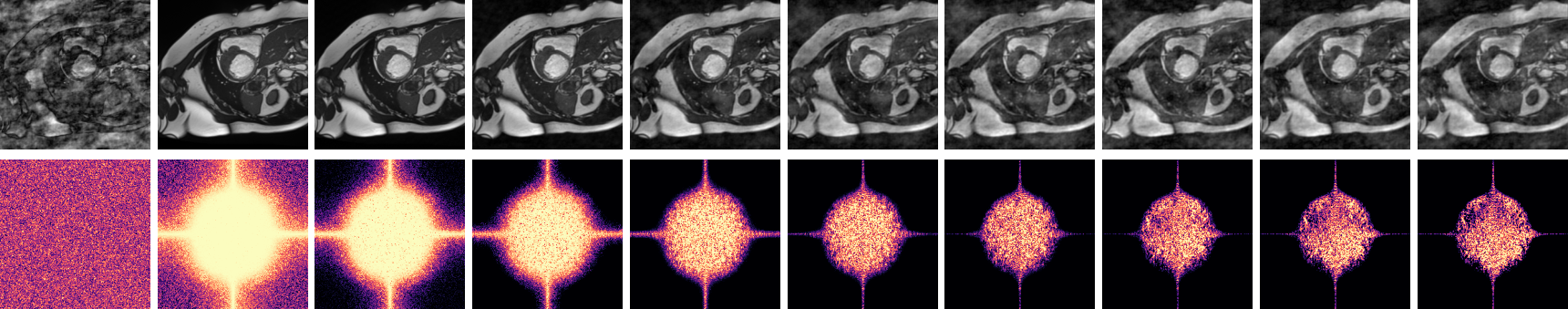}};
        \node [right=of frame1.east, xshift=0.5mm] (frame2)
            {\includegraphics[width=0.095\textwidth]{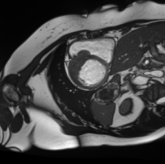}};
        
        \node [right=of frame1.east, yshift=14mm, xshift=-1.5mm] (a1) {};
        \node [right=of frame1.east, yshift=-14mm, xshift=-1.5mm] (a2) {};
        \draw [thick] (a1) -- (a2);
        
        \node [above=of frame1.west, yshift=15mm, xshift=2mm, align=center] (text1) {};
        \node [above=of frame1.east, yshift=15mm, xshift=-2mm, align=center] (text2) {};
        
        \draw [->] (text1) -- node [above,midway] {{\tiny \berm\ undersampling mask  optimization}} (text2);

    \end{tikzpicture}
    \caption{
    Visualization of the \berm\ optimization routine for the ACDC dataset with an acceleration factor of \texttt{x8} for a 2D mask.
    The \textbf{bottom row} shows our Bernoulli mask distribution $\post$, where a lighter color implies a higher probability for sampling the respective entry in the cartesian k-space grid.
    Starting from a randomly initialized distribution,
    \berm\ gradually optimizes $\post$ to maximize reconstruction quality while simultaneously increasing the sparsity of the masks.
    The resulting distribution converges to a domain-specific mask with desired acceleration factor that preserves most of the image's quality (\textbf{top row}). 
    The original image is displayed on the \textbf{right}.
    }
    \label{fig:intro}
\end{figure*}
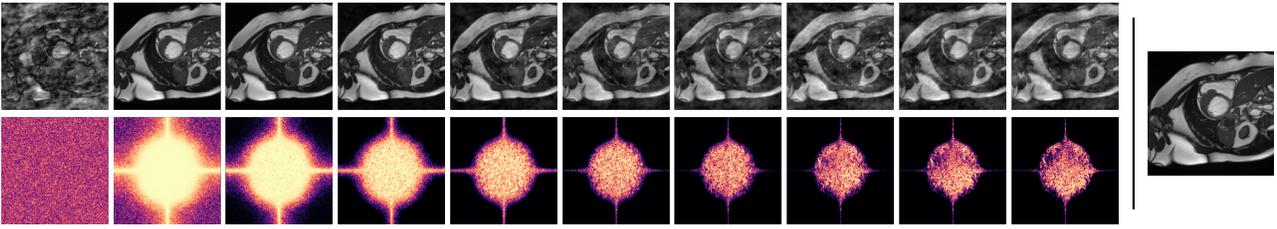

In contrast to previous work, we examine the potential of directly learning the mask itself via projected gradient descent.
Our method called \berm\ is a fully differentiable, learning-based, and \textbf{pro}babilistic framework for \textbf{m}ask optimization.
By framing the search for an optimal mask as a probabilistic optimization problem for a pre-specified acceleration factor, \berm\ is able to find the optimal undersampling distribution using ideas from relaxed categorical optimization in deep learning research.
\berm\ optimizes the parameters of the undersampling distribution directly without the need for a model, resulting in runtimes of up to a few seconds.
\berm\ does not operate under the assumption of any predetermined undersampling pattern.
As such, it adapts to deliver bespoke results tailored to the specific requirements of the downstream task and the anatomical region as depicted in Figure~\ref{fig:intro}.
This adaptability makes \berm\ a versatile, data-driven tool, capable of serving as an all-purpose mask generator.

\section{Related Work}

The field of learning-based undersampled MRI reconstruction can largely be divided into two main branches: one focused on enhancing the quality of the reconstructed images, and the other on optimizing the initial sampling pattern itself.

\paragraph{Image enhancement.}
In the field of image quality enhancement, \cite{gao_projection-based_2022} consider the auto-regressive nature of encoder-decoder transformer architectures to derive missing spokes in radial undersampling.
Using adversarial methods, \cite{belov_towards_2021} is able to enhance images with extreme acceleration factors. 
\cite{chung_score-based_2022} apply the conditioning mechanism in diffusion models to retrieve enhanced images from low-quality reconstructions.
Instead of conditioning on the actual image, \cite{peng_towards_2022} utilize the undersampled mask directly to guide the diffusion process.
To mitigate domain shifts, \cite{gungor_adaptive_2022} propose a diffusion prior trained via adversarial mapping.
Through open-sourcing high-quality and large MRI datasets, the fastMRI \cite{zbontar_fastmri_2019} challenge actively promotes development in the area of enhancing undersampled MRI.

\paragraph{Mask optimization and hybrid approaches.}
While most approaches that optimize masks fall under the category of hybrid approaches (see below) as they also try to enhance the image quality at the same time, a few learning-based optimization approaches exist that exclusively focus on the sampling pattern.
Iterative gradient sampling (IGS; \cite{razumov_optimal_2022, razumov_optimal_2022-1}) is capable of efficiently deriving undersampling masks in 1D, but its complexity scales quadratically when moving to a 2D setting, whereas \berm's complexity is only dependent on the number of optimization steps.
\cite{bahadir_learning-based_2019} propose LOUPE which encourages sparsity of the mask through $\ell_1$-penalization (without directly enforcing a specific acceleration factor) and approximates mask probabilities using a sigmoid output.
As this approach includes a reconstruction network, it can also be seen as a hybrid approach.
A different network-assisted idea is to employ a binarized mask on the forward pass and -- in contrast to our proposal -- transfer its gradients to a continuous real-valued mask on the backward pass \cite{weiss_joint_2020, xue20222d}.
An updated binary mask with induced sparsity is then produced on the subsequent iteration by thresholding and pruning the real-valued scores.
With a focus on enforcing physical constraints, \cite{lazarus2019sparkling} propose optimized trajectories for compressed sensing.
Similarly, \cite{weiss2019pilot} solve for k-space trajectories, but additionally use a reconstruction network to enhance image quality.
\cite{wang_b-spline_2022} jointly optimize quadratic B-spline kernels together with a reconstruction network for multi-coil data.

\paragraph{Other approaches.} Further approaches include variational models to jointly synthesize and reconstruct MRI images \cite{bian_learnable_2022}, sharpening networks \cite{dong_invertible_2022} to counter the absence of high-frequency features in undersampled MRIs, or learnable Fourier interpolation \cite{ding_mri_2022}.
To account for meta information such as the manufacturer, \cite{liu_undersampled_2022} condition the reconstruction network on side information.

\section{Methods}

In our presentation on learning a distribution for fully probabilistic undersampling masks, we first describe our routine for a single image and later extend this idea to jointly optimize masks across a whole dataset.

\subsection{Probabilistic Undersampling Masks}

In the following, we define $D$ as the number of elements on the 2-dimensional or 3-dimensional k-space grid and use a vectorized notation for all objects for simplicity.
Thus, $\x_k = (x_k^{(1)}, \ldots, x_k^{(D)})^\top \in \mathbb{C}^{D}$ depicts an image residing in k-space.
Partial acquisition is augmented by applying a binary mask $\m \in \{0, 1\}^D$ to the fully-sampled $\x_k$ element-wise: $\x_k \odot \m$. For a given $\m$, the translation of the image from the under-sampled k-space to the complex image domain $\mathcal{X} \subseteq \mathbb{C}^{D}$ is done by $\x_c = \mathcal{F}^{-1} (\x_k \odot \m)$, where $\mathcal{F}^{-1}$ is the inverse Fourier transform matrix. This procedure amounts to a simple zero-filling reconstruction strategy.

\paragraph{Empirical Bayes Optimization.} Instead of following a distinct sampling pattern for \m, we assume that every element $m^{(i)}, i = 1, ..., D,$ in $\m$ is the result of an independent Bernoulli experiment of a random variable $M^{(i)}$ with distribution defined by 
\begin{equation} \label{eq:prior}
   \prior := \mathbb{P}(\bm{M} = \m \mid \s) = \prod_{i=1}^D \theta_i^{m^{(i)}} (1-\theta_i)^{1-m^{(i)}},
\end{equation}
$\bm{M} = (M^{(1)},\ldots,M^{(D)})^\top$, and $\theta_i \in (0, 1)$ the sampling probability of the $i$th element.
Eq.~\ref{eq:prior} can be thought of as a Bayesian prior for the image mask. While this factorized prior allows the method to be adaptable for any task and domain, other choices are possible and discussed in more detail in Section~\ref{sec:conclusion}.

Given the prior distribution, we strive to optimize the posterior sampling distribution of masks (i.e., after accounting for the specific reconstruction task and data) for arbitrary differentiable loss functions $\mathcal{L}$, in particular those used in computer vision.
As these loss functions are typically designed to only work in real-valued space, we transform $\x_c$ into a  real-valued representation using its magnitude image $\hat{\x} = | \x_c | \in \mathbb{R}^D$. 
The quality of the reconstruction can then be assessed by $\mathcal{L}(\hat{\x}, \x)$, where $\x$ is the fully-sampled original image and $\hat\x$ the undersampled image as a function of $\m$ (or a random variable when $\M$ is not yet observed).
Given a data point $\x$, we take an empirical Bayes approach and directly optimize \s\ by minimizing the expected loss via
\begin{equation} \label{eq:obj}
       \argmin_{\s} \mathbb{E}_{\bm M        \sim \post} \mathcal{L}(\hat{\x}, \x)  
        \approx \argmin_{\s} \frac{1}{L} \sum_{l=1}^L \mathcal{L}(\hat{\x}^{(l)}, \x) \eqendc 
\end{equation}
where we approximate the expectation w.r.t.~the posterior distribution $\post$ using $L$ Monte-Carlo samples $\hat{\x}^{(l)}$ from $\post$.
Although in general, no analytical representation of the posterior exists, the optimization of Eq.~\ref{eq:prior} and $\frac{1}{L} \sum_{l=1}^L \mathcal{L}(\hat{\x}^{(l)}, \x)$ individually is straightforward and allows us to derive an iterative optimization procedure (Algorithm~\ref{alg:berm}) to optimize $\s$.

\begin{figure}[htbp]
    \centering
    \includegraphics[width=\linewidth]{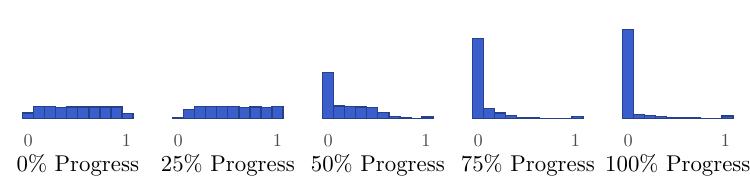}
    \caption{Histograms of probabilities in \s\ during different optimization phases for a sample of fastMRI Knee. Starting from a random initialization (leftmost distribution), the sparsification constraint leads to a posterior $\post$ with probabilities that tend to be close to zero or one.}
    \label{fig:optimization-hist}
\end{figure}

\paragraph{Constrained Optimization.} Without further constraints, the optimal solution of Eq.~\ref{eq:obj} is $\s = \bm 1$, i.e., the fully-sampled image with minimal loss, as $\hat{\x} \equiv \x$.
To incorporate undersampling into our objective, we introduce a constraint similar to \cite{zhou_effective_2021} by limiting the sum of all probabilities in $\post$ to a pre-specified value $S$, i.e., $\sum_{i=1}^D \theta_i \leq S$.
Practically speaking, $S$ will result in the number of sampled k-space elements as $\sum_{i=1}^D \theta_i$ is the expected value of $\lvert\lvert\bm m\rvert\rvert_0$.
Given a user-defined acceleration factor $\alpha$, i.e. the ratio between the amount of acquired k-space points in the full versus the undersampled image, we can alternatively define $S = \lfloor \frac{D}{\alpha} \rfloor$ and our final objective as
\begin{equation} \label{eq:constraint}
\begin{gathered}
    \argmin_{\s} \mathbb{E}_{\bm M \sim \post} \mathcal{L}(\hat{\x}, \x) \\
    s.t. \:  \textstyle \sum_{i=1}^D \theta_i \leq S \: \text{and} \: S \in \{0, ..., D\} \eqendp
\end{gathered}
\end{equation}
The constraint in Eq.~\ref{eq:constraint}, which can equally be expressed as an $\ell_1$-norm penalty for \s, has an intrinsic affinity for sparse mask distributions (c.f.~Figure~\ref{fig:optimization-hist}).

\subsection{Differentiability through Reparameterization}

Except for the stochastic sampling of \m, the proposed approach is fully differentiable including $\mathcal{F}^{-1}$ and $|.|$. 
In order to use modern autograd frameworks for stochastic masks, we apply the Gumbel-Softmax trick  \cite{jang_categorical_2017} tailored to the Bernoulli distribution \cite{zhou_effective_2021}. Let $\bm{\rho} := \log \left( \frac{\s}{1 - \s} \right)$ be the log odds-ratio for $\s$. Then, a ``soft mask'' $\bm m_{\text{\scriptsize soft}}$, allowing for differentiability, can be obtained by sampling
\begin{equation} \label{eq:approxsampling}
\begin{split}
            \bm m_{\text{\scriptsize soft}} & \sim \mathds{1} \left( \bm{\rho} + \bm g_1 - \bm g_0 \geq \bm 0 \right) \\ 
            & \approx \sigma \left((\bm{\rho} + \bm g_1 - \bm g_0) \tau^{-1}\right) \eqendc
\end{split}
\end{equation}
where $\bm g_1, \bm g_0$ are independent and identically distributed samples from a Gumbel(0,1)~distribution and the indicator function $\mathds{1}$ is relaxed using a sigmoid function $\sigma$.
A temperature parameter $\tau$ controls the \textit{softness} of the discrete approximation and is annealed during optimization.
Here, stochasticity is rerouted over the Gumbel samples and thus a computational graph is able to propagate gradients to \s.
As the mask $\bm{m}$ needs to be strictly binary, which is not the case for $\bm m_{\text{\scriptsize soft}}$, we adopt the straight-through Gumbel estimator trick \cite{jang_categorical_2017}, yielding 
\begin{equation}
    \m =  \mathds{1} \left( \bm m_{\text{\scriptsize soft}} \geq 0.5 \right) - sg \lbrack \bm m_{\text{\scriptsize soft}} \rbrack + \bm m_{\text{\scriptsize soft}} \eqendc
    \label{eq:straight}
\end{equation}
where $\mathds{1}$ is applied element-wise and returns our final binary mask sample.
$sg$ denotes the \textit{stop gradient} operation, which blocks gradients from backpropagation.
In other words, Eq.~\ref{eq:straight} yields discrete values while we obtain gradients for its soft approximation.

\subsection{Optimization via Projected Gradient Descent}

\begin{figure}[htbp]
    \centering
    \begin{tikzpicture}[
         image/.style = {text width=0.9\linewidth, 
                         inner sep=0pt, outer sep=0pt},
        node distance = 1mm and 1mm
                                ] 
        \node [] (frame1)
            {\includegraphics[width=0.95\linewidth]{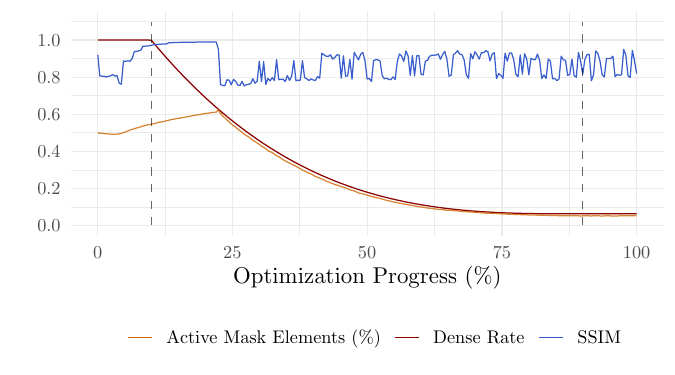}};

        \node [above=of frame1, yshift=-4mm, xshift=3mm] (a1) {\small \textbf{Constraining}};
        \node [left=of a1, xshift=-9mm] (a2) {\small \textbf{Exploration}};
        \node [right=of a1, xshift=8mm] (a3) {\small \textbf{Exploitation}};
        
    \end{tikzpicture}
    \vspace{-5mm}
    \caption{Sparsity versus reconstruction quality for a sample of fastMRI Knee. The amount of active masked elements in $\m$ is bounded by the continuously annealed dense rate $\frac{S}{D}$. During optimization, the average SSIM metric (measuring reconstruction) plummets when \s\ is limited by the constraint (at around 20\% progress) but roughly stays constant or even slightly improves while the number of active elements is further restricted.}
    \label{fig:optimization-progress}
\end{figure}
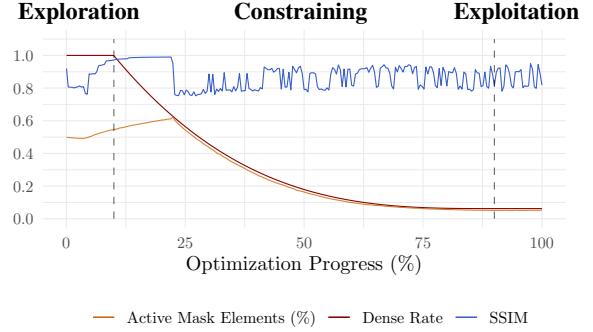

The optimization problem in Eq.~\ref{eq:constraint} cannot be solved effectively with standard gradient descent.
Instead, we follow \cite{zhou_effective_2021} and use a projected gradient approach by first updating the unconstrained parameter vector $\tilde{\s} = \lbrack \tilde{\theta}_1, ..., \tilde{\theta}_D \rbrack$ using $\tilde{\s}  = \s - \eta \nabla_{\s}\mathbb{E}_{\bm M \sim \post} \mathcal{L}(\hat{\x}, \x)$ with $\eta$ being the learning rate, and then project $\tilde{\s}$ into the space of valid elements by solving
\begin{equation}
    \textstyle \sum_{i=1}^D \min (1, \max(0, \tilde{\theta_i} - \lambda)) = S
    \label{eq:projdesc}
\end{equation}
for $\lambda \in \mathbb{R}$, yielding
\begin{equation}
    \s = \min (\bm 1, \max (\bm 0, \tilde{\s} - \max(0, \lambda) \bm 1)) \eqendp
    \label{eq:projeq}
\end{equation}
See Supplementary Material~\ref{app:proj} for an extended derivation of the objective.
A solution to Eq.~\ref{eq:projdesc} can be obtained using a convex solver or root-finding method such as bisection search.
To foster exploration at the beginning of the training and allow for exploitation at later stages, we anneal $S$ during optimization.
First, exploration iterations with $S = D$ allow for unrestricted optimization.
Then, an annealing phase following the schedule of \cite{zhu_prune_2017} (see Supplementary Material~\ref{app:anneal}) decreases $S$ to meet the desired acceleration factor.
Finally, the exploitation phase optimizes \s\ under the nominal constrained $S$ (c.f.~Figure~\ref{fig:optimization-progress}).

Our procedure is summarized in Algorithm~\ref{alg:berm} for a single image $\x_k$.
While our previous approach is presented for only a single sample, it can be equivalently applied to a whole dataset by iteratively taking random batches instead of a single $\x_k$ for the optimization of $\bm{\s}$.
Note that the only trainable parameters in \berm\ are the $D$ parameters $\bm{\s}$.
Optimization can thus be done in a matter of seconds (single image) or a few minutes (full dataset).
\begin{algorithm}
\renewcommand{\algorithmicrequire}{\textbf{Input:}}
\caption{Optimization routine of \berm}\label{alg:berm}
\begin{algorithmic}
\Require k-space image $\x_k$, mask distribution $\prior$, number of samples $L$, acc. factor $\alpha$, iterations $i$, learning rate $\eta$, criterion $\mathcal{L}$

\For{$1, ..., i$}
    \State \Comment{Draw samples (cf. Eq. (\ref{eq:approxsampling}) and (\ref{eq:straight}))}
    \State $ \mathcal{M} \gets \{ \m_1, ..., \m_L \}$ 
    \State \Comment{Expand image to match shape of $\mathcal{M}$}
    \State $\bm X_k \gets \text{expand}(\x_k)$ 
    \State \Comment{Compute undersampled magnitude images}
    \State $\hat{\bm X} \gets \lvert \mathcal{F}^{-1} (\bm X_k \odot \mathcal{M}) \rvert$
    \State \Comment{Compute fully-sampled image and expand}
    \State $\bm X \gets \text{expand}(\lvert \mathcal{F}^{-1} (\x_k) \rvert)$ 
    \State \Comment{Compute batched loss and apply gradients}
    \State $\tilde{\s}  \gets \s - \eta \nabla_{\s}\mathcal{L}(\hat{\bm X}, \bm X)$ 
    \State \Comment{Anneal to acc. factor over iterations}
    \State $S \gets \text{anneal}(\alpha)$ 
    \State \Comment{Project updated weights (cf. Eq. (\ref{eq:projdesc}) and (\ref{eq:projeq}))}
    \State $\s \gets \text{project}(\tilde{\s}, S)$ 
\EndFor
\end{algorithmic}
\end{algorithm}

\subsection{Constraint Posterior Mode} \label{sec:postmode}
\begin{figure*}
    \centering
    \resizebox{0.8\textwidth}{!}{
    \begin{tabular}{lcccc@{\hskip 0.5cm}|@{\hskip 0.5cm}c}
    & \texttt{x4} & \texttt{x8} & \texttt{x16} & \texttt{x32} & \textbf{Example} \\
    
    \tikz \node[rotate=90,]{\hspace{5mm} \textbf{ACDC}}; & \includegraphics[width=0.20\textwidth]{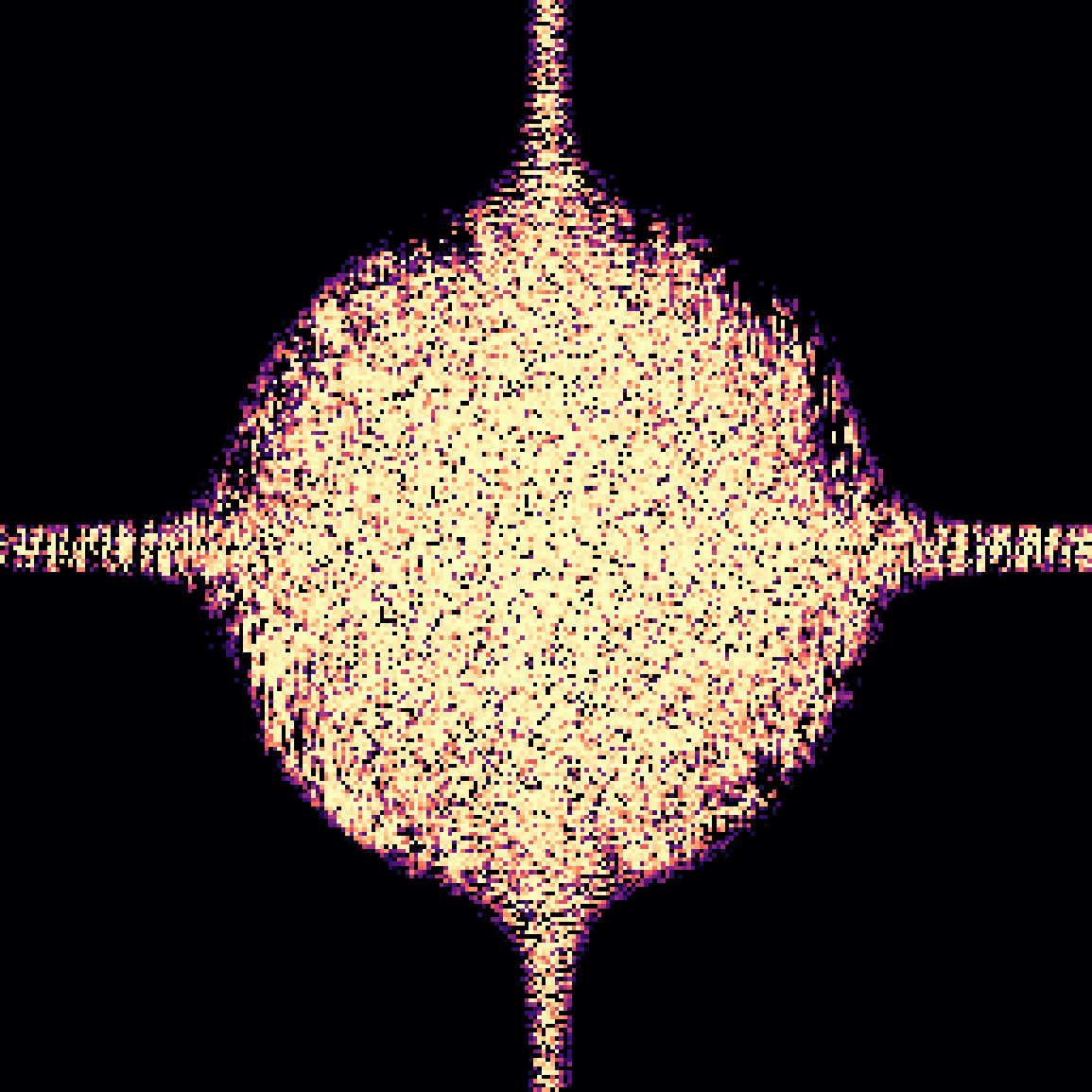} &   \includegraphics[width=0.20\textwidth]{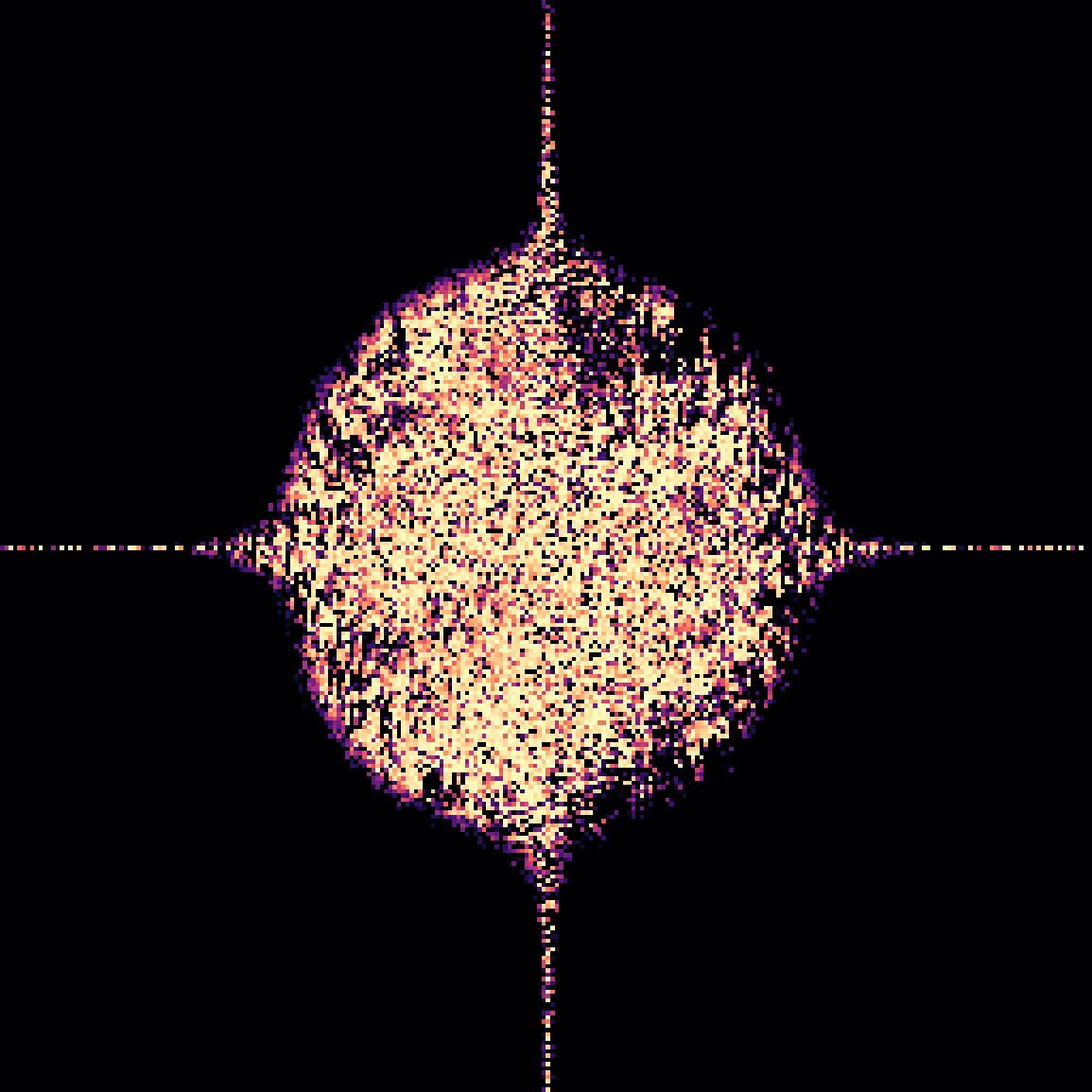} &
  \includegraphics[width=0.20\textwidth]{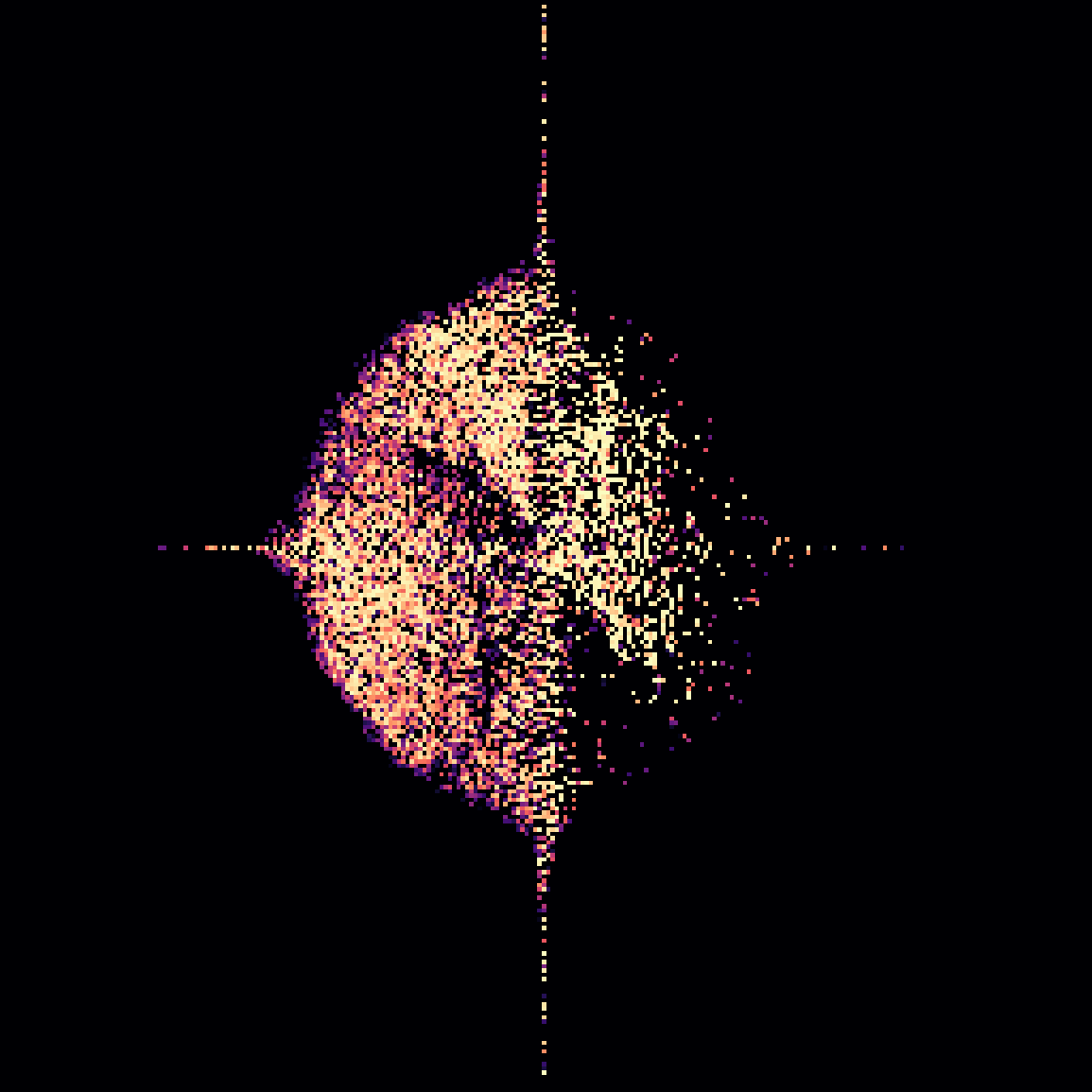} &   \includegraphics[width=0.20\textwidth]{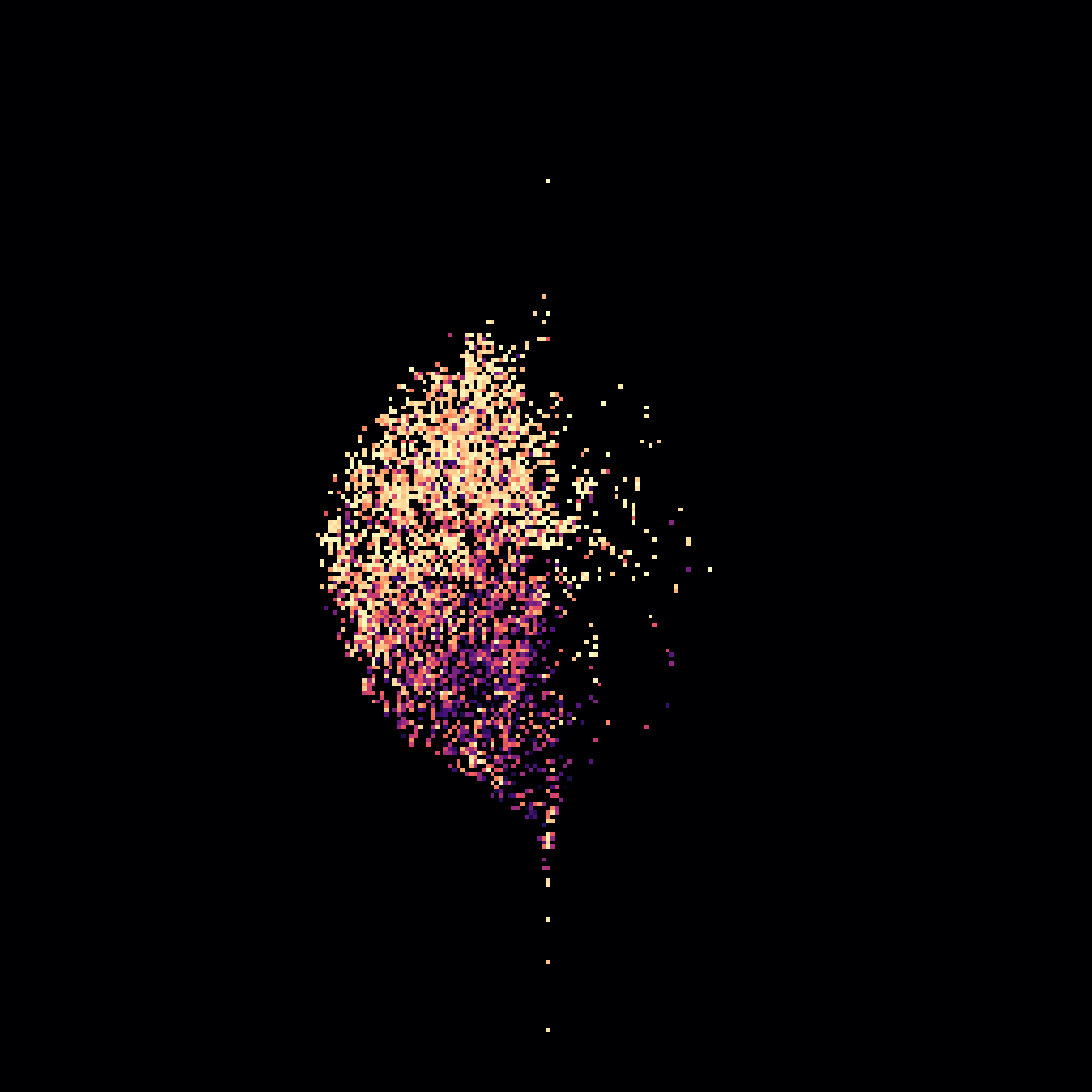} &
  \includegraphics[width=0.20\textwidth]{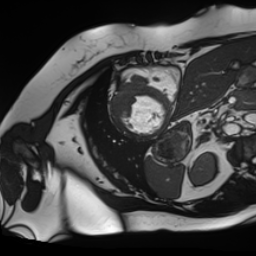}
  \\
  
    \tikz \node[rotate=90]{\hspace{5mm} \textbf{BraTS}}; & \includegraphics[width=0.20\textwidth]{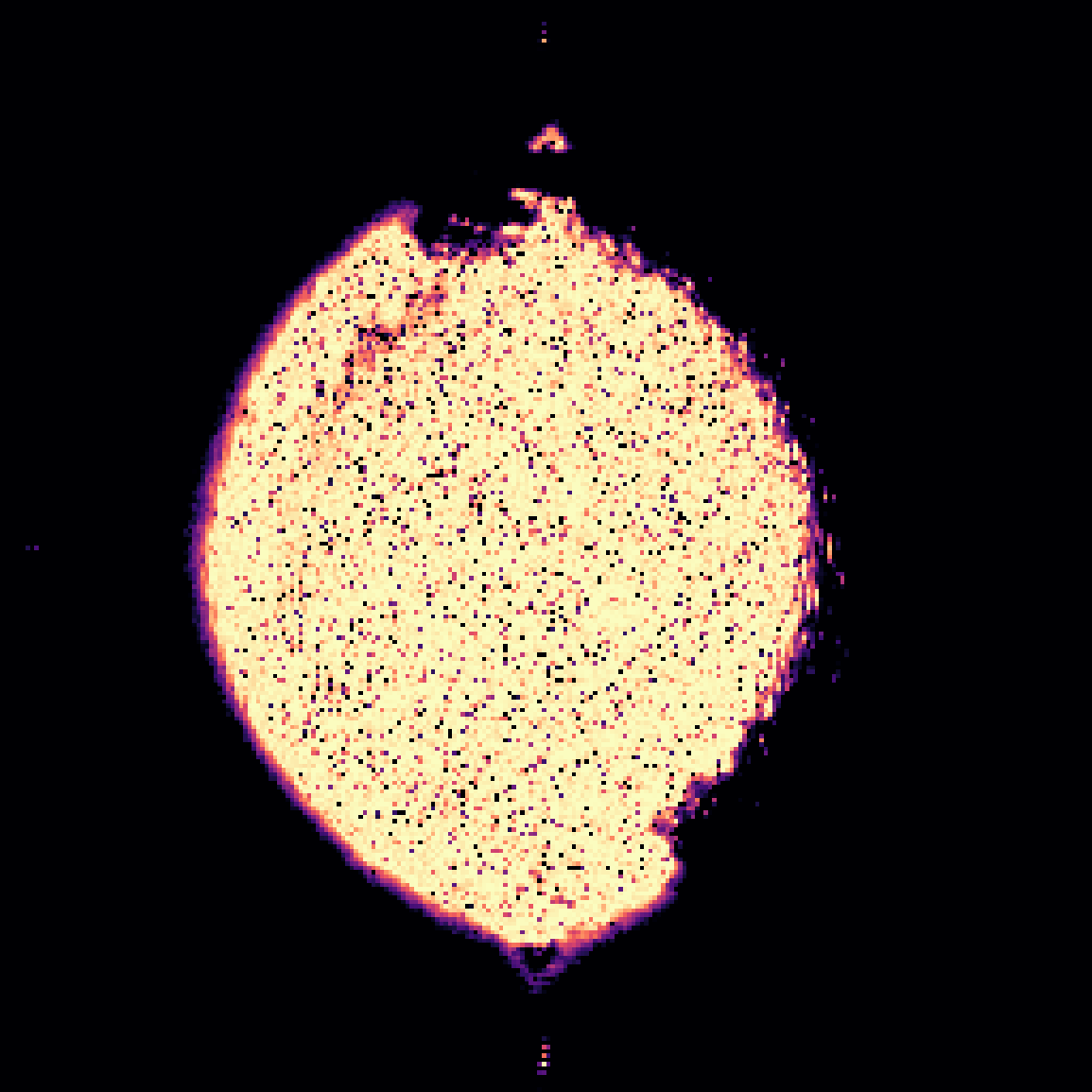} &   \includegraphics[width=0.20\textwidth]{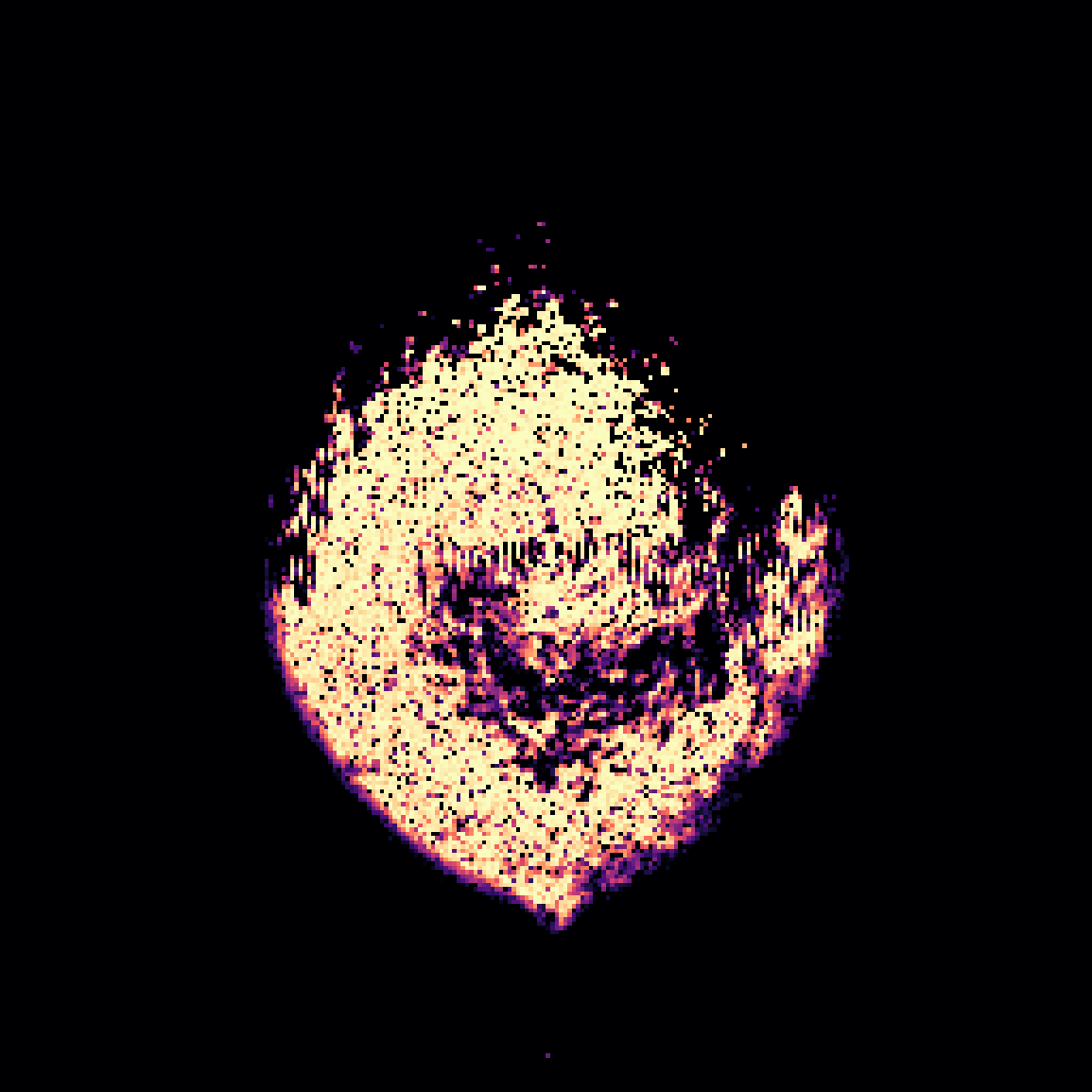} &
  \includegraphics[width=0.20\textwidth]{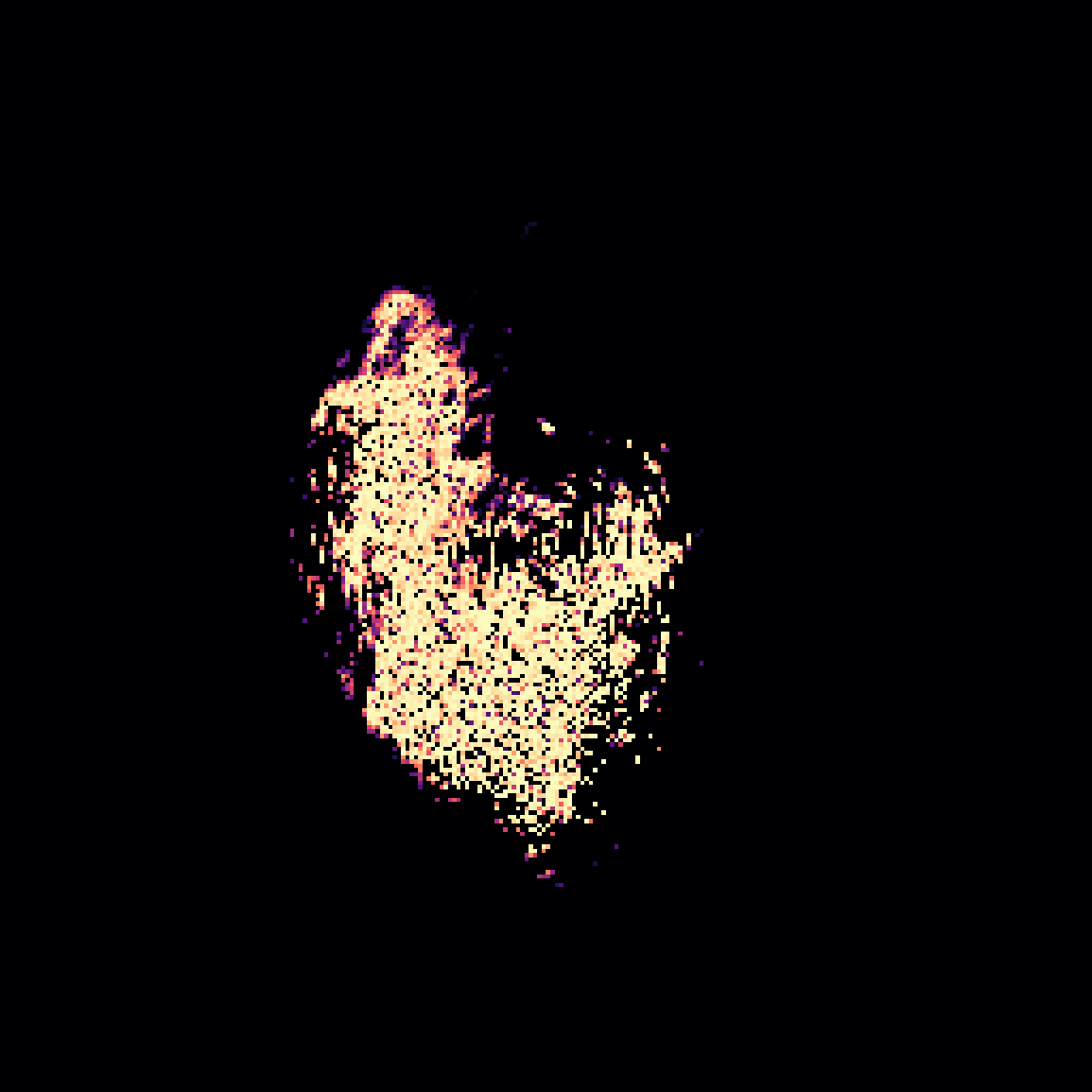} &   \includegraphics[width=0.20\textwidth]{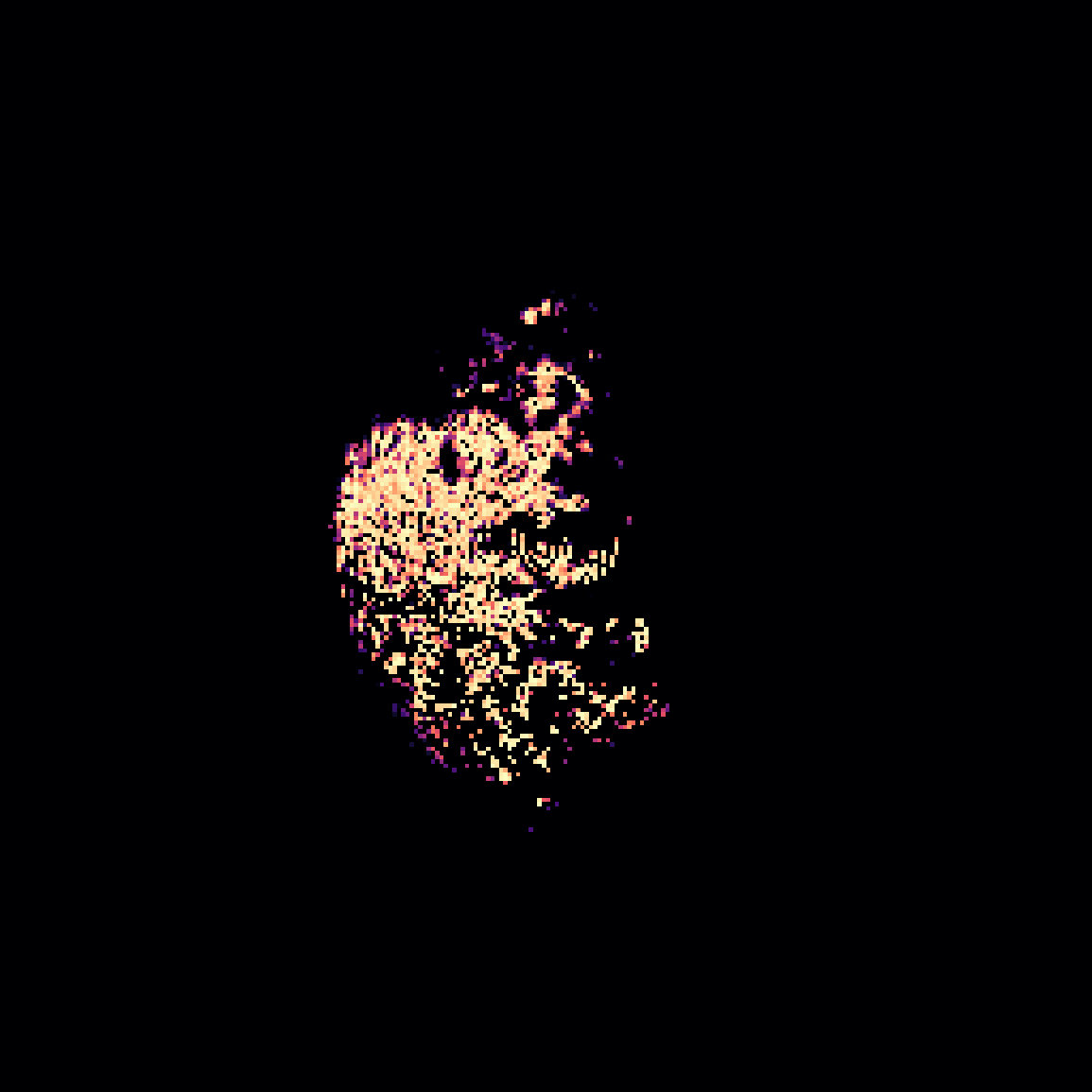} &
  \includegraphics[trim={11mm 11mm 11mm 11mm},clip, width=0.20\textwidth]{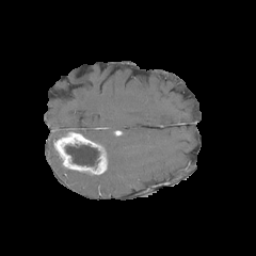}
  \\
  
     \tikz \node[rotate=90]{\hspace{-2.5mm} \textbf{fastMRI Knee}}; & \includegraphics[width=0.20\textwidth]{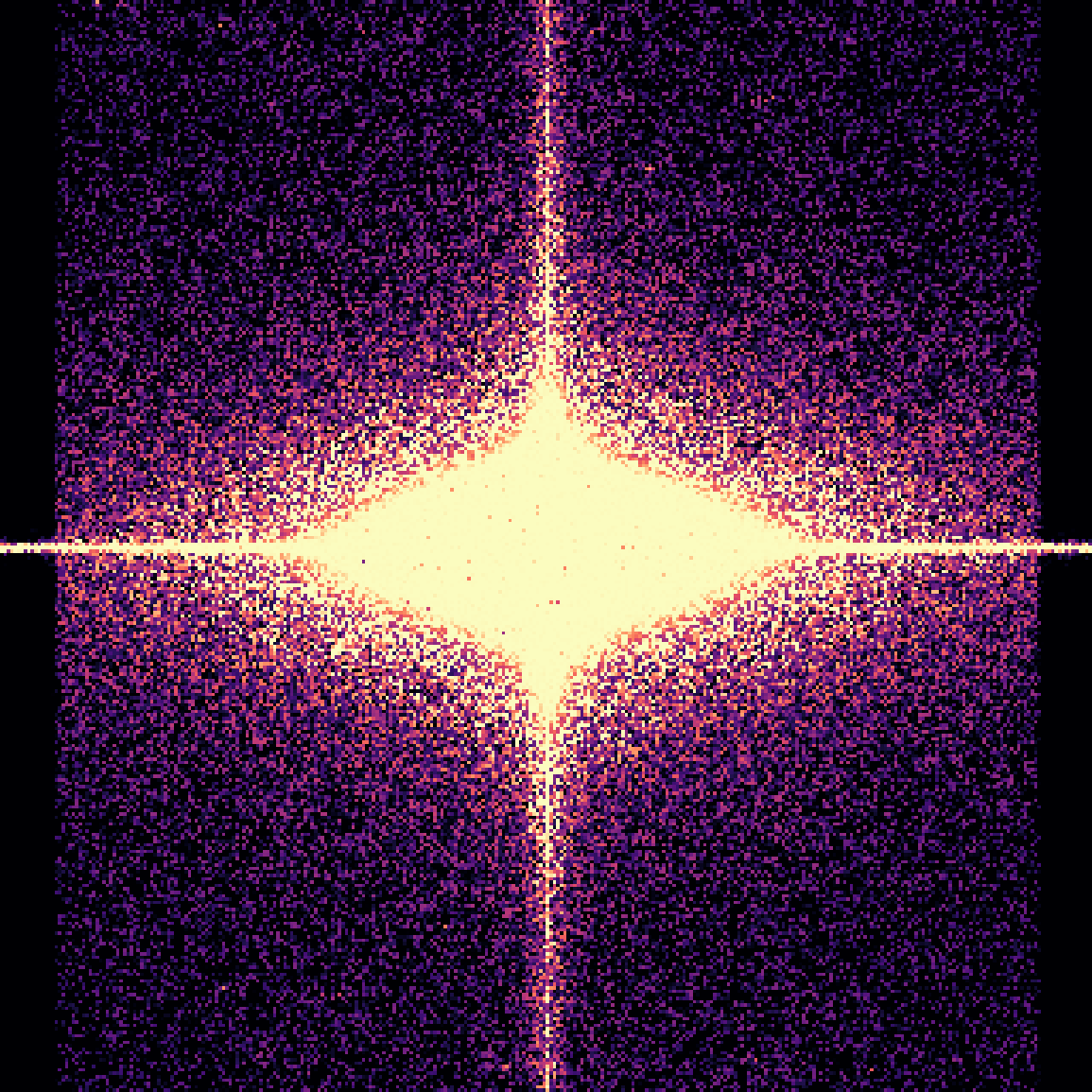} &   \includegraphics[width=0.20\textwidth]{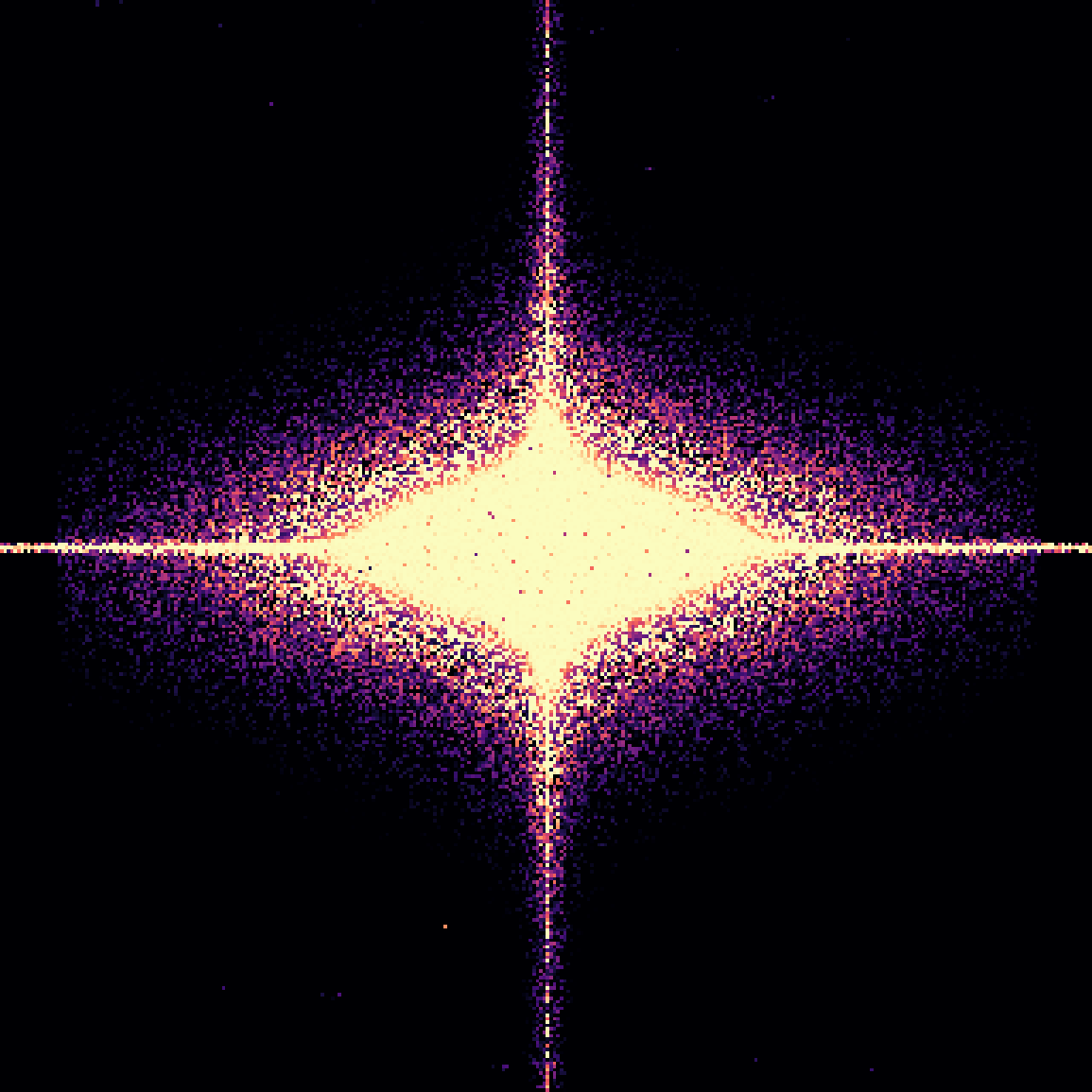} &
  \includegraphics[width=0.20\textwidth]{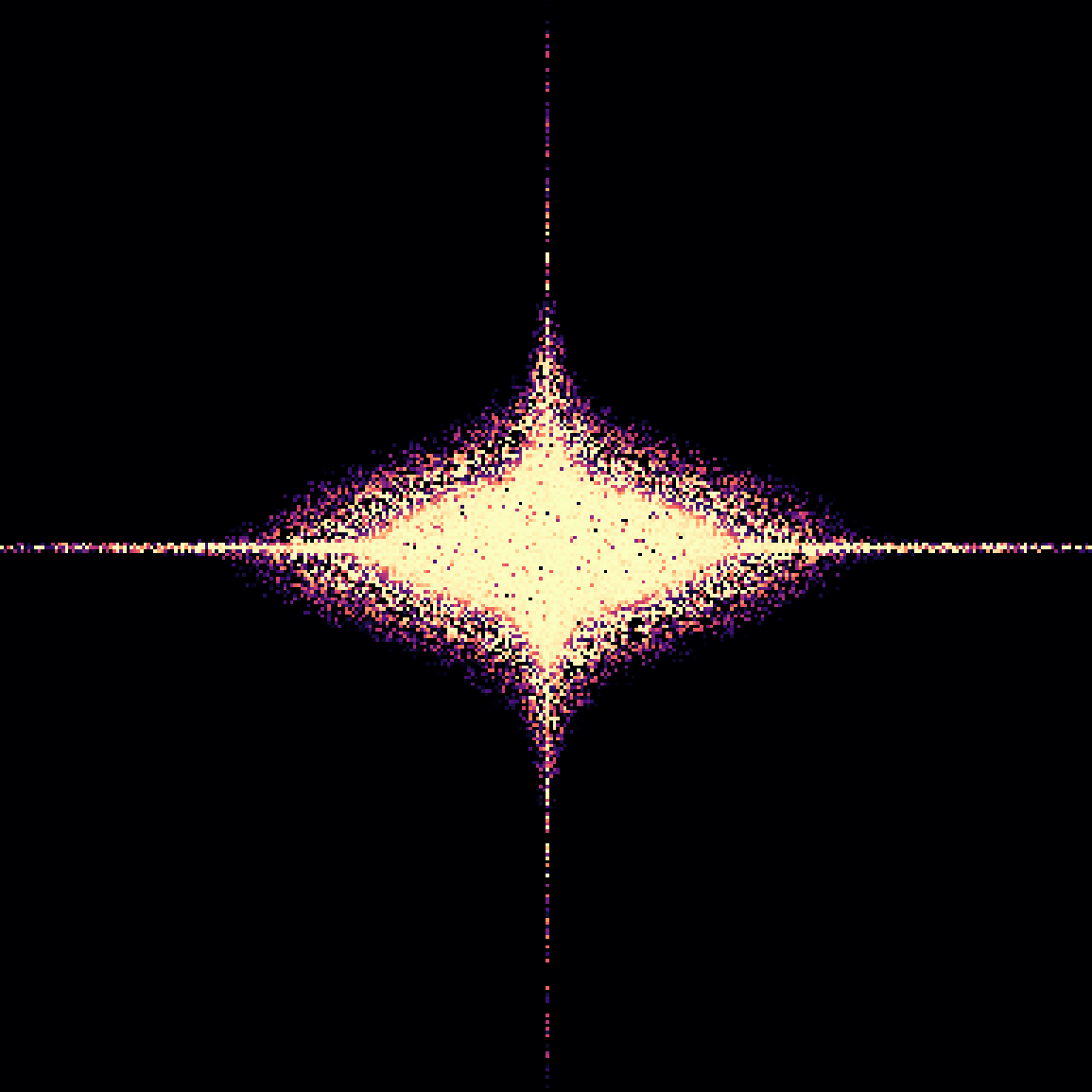} &   \includegraphics[width=0.20\textwidth]{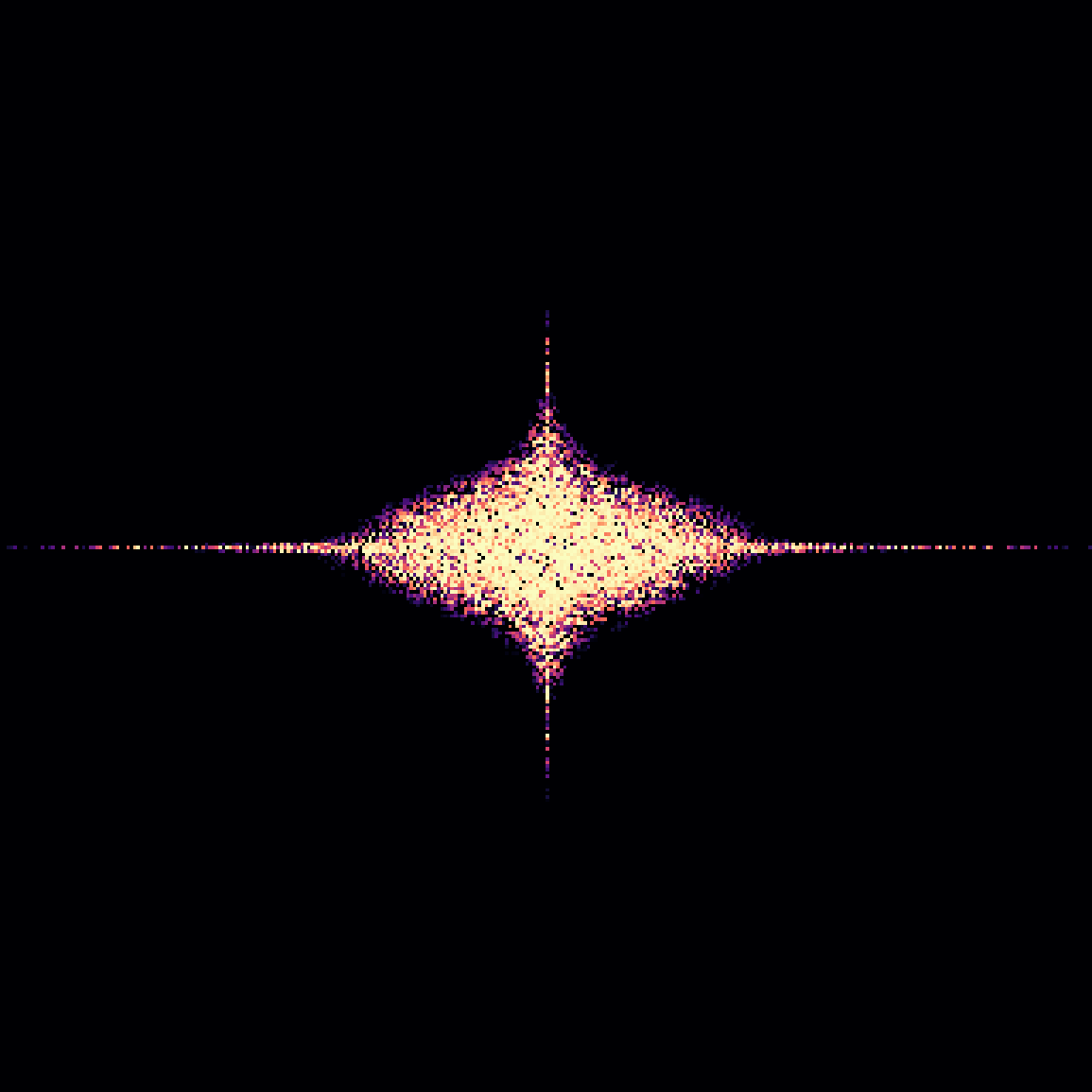} &
  \includegraphics[width=0.20\textwidth]{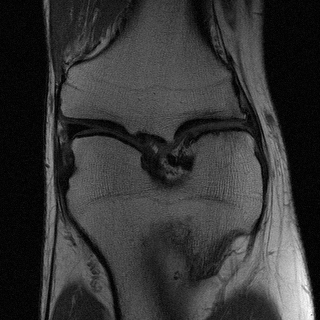}
  \\
    \end{tabular}
    }
    \caption{Optimized undersampling 2D mask distributions for ACDC, BraTS and fastMRI Knee (\textbf{rows}) with varying acceleration factors (\textbf{columns}). Different anatomic regions (\textbf{right}) have a distinct unique optimal distribution. The color indicates the value of each $\theta_i$ representing a point in k-space. Bright color implies a high probability of acquiring the respective entry.}
    \label{fig:ex-acc_fac}
\end{figure*}
Although realizations of the resulting distribution $\post$ of our approach are sparse in practice (cf.~Figure~\ref{fig:optimization-hist}) and adhere to the constraint, no real data set will result in a posterior with all $\theta_i$ values being optimized to be either exactly 0 or 1.
Whereas this result simply reflects the uncertainty in the data, some practical applications require masks to be deterministic.
Therefore, we propose a deterministic variant of our approach by generating the mask $\m$ using the mode of (the optimized) posterior $\post$ conditional on the constraint 
$\lvert\lvert\bm m\rvert\rvert_0 = S$.
In other words, we find this mask $\m^*$ as
\begin{equation}
   \m^* = \argmax_{\bm m, \lvert\lvert\bm m\rvert\rvert_0 = S} \post(\bm m) \eqendp
\end{equation}
This approach is equivalent to setting all mask entries ${m^*}^{(i)}$ corresponding to the $S$ largest probabilities $\theta_{(D-S+1)},\ldots,\theta_{(D)}$ to the value one and all others to zero.

\section{Experiments}

We investigate the performance of \berm\ using slices from ACDC \cite{bernard_deep_2018}, BraTS \cite{menze_multimodal_2015, bakas_advancing_2017, bakas_identifying_2019} and fastMRI Knee \cite{zbontar_fastmri_2019}:

\textbf{ACDC} are cardiac MRIs with 100 train and 50 test subjects.
We extract the end-diastolic frame in 256px resolution including segmentation labels of the left and right ventricular cavity as well as the left ventricular myocardium, yielding $548$ train and $338$ test slices.
k-space data is emulated via Fourier transform. 

\textbf{BraTS} contains brain MRIs with T2-FLAIR, T1-, T1Gd- and T2-weighted modalities.
The goal of the segmentation is to determine the classes of whole, core, and enhancing tumors.
The dataset is split into 387 train and 97 test subjects. We extract $19,350$ train and $4,850$ test slices using slice indices between $60$ and $110$ with 256px resolution. 
k-space data is emulated as in ACDC. 

\textbf{fastMRI Knee} includes raw k-space data of single coil knee MRI. To focus on pathologies, we extract the annotated subset of fastMRI+ \cite{zhao_fastmri_2021} amounting to $8,057$ train and $1524$ test slices with a center-crop to 320px resolution.

For \berm\ we use 2500 iterations with a learning rate of $0.01$ in the Adam \cite{kingma_adam_2017} optimizer.
From the full amount of iterations, we took 250 steps for exploration and exploitation each.
This configuration provided an ideal balance between runtime and convergence.
We use batches of size 32 and draw $L=4$ Bernoulli samples for each sample, yielding a total batch size of 128.
Similar to \cite{kingma_auto-encoding_2022}, we found that a low number of Monte Carlo samples is sufficient if the batch size is large.
The temperature $\tau$ follows a linearly decreasing schedule from $1$ to $0.03$ in the last step, which aligns with \cite{zhou_effective_2021}.
For reconstruction, we use the mean squared error for $\mathcal{L}$.
The mask distribution $\prior$ is initialized by drawing $\theta_i$ from $U(0,1)$.
The final masks are obtained by applying Bayesian model averaging on 10 different \berm\ solutions.

Optimization is done in PyTorch v.1.13 \cite{paszke_pytorch_2019} on an NVIDIA A100 GPU.
We compare our approach against an equispaced mask with fully-sampled central region of 4\% \cite{zbontar_fastmri_2019} and random offset, a 2D variable density Gaussian mask, and the learning-based IGS \cite{razumov_optimal_2022-1}) method.
The reported metrics represent the mean across 10 randomly initialized runs per method.

\subsection{Domain-specific Masks}

Each anatomic region (as a composition of different elemental shapes) yields a distinct k-space representation.
Our first experiment not only demonstrates that this is the case in practice but also shows that a data-driven optimization routine for masks such as \berm\ is indeed necessary to facilitate optimal reconstruction (c.f.~Figure~\ref{fig:ex-acc_fac} for the results of \berm\ trained on the three different datasets).
For instance, the cardiac ACDC dataset consists dominantly of elliptic primitives, which results in a completely different optimal mask than fastMRI Knee with a lot of vertical lines and some horizontal elements in image space.

\subsection{Convergence and Stability}
\begin{figure}[b]
    \centering
    \addtolength{\tabcolsep}{-3pt}  
    \resizebox{\linewidth}{!}{
    \begin{tabular}{cccc}
        \texttt{x4} & \texttt{x8} & \texttt{x16} & \texttt{x32} \\
         \includegraphics[width=0.3\linewidth]{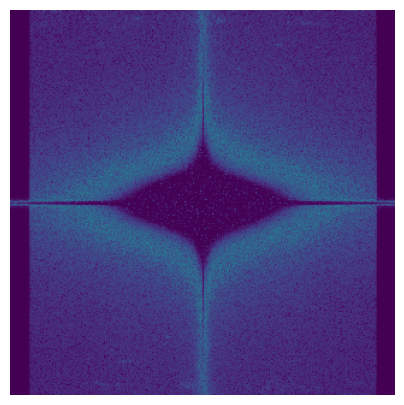} &  
         \includegraphics[width=0.3\linewidth]{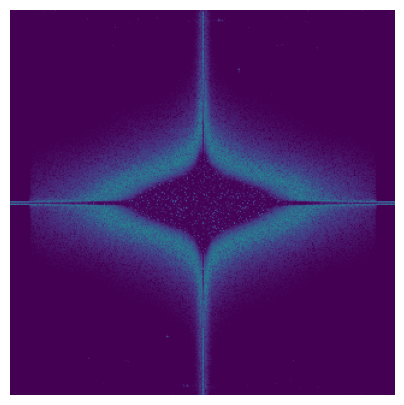} &  
         \includegraphics[width=0.3\linewidth]{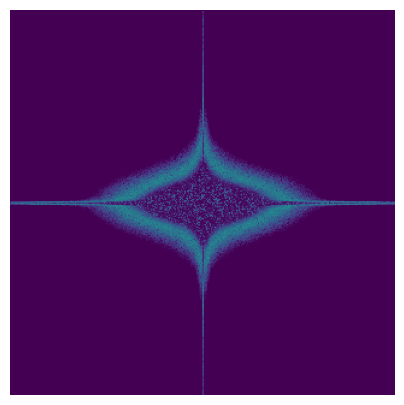} &  
         \includegraphics[width=0.3\linewidth]{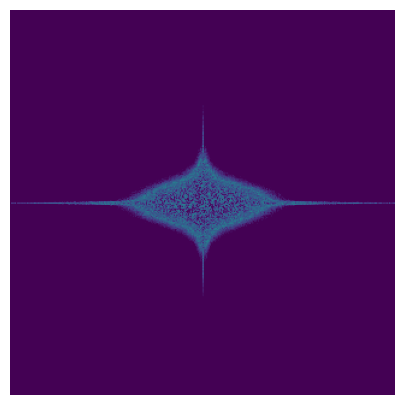}
    \end{tabular}
    }
    \addtolength{\tabcolsep}{3pt}  
    \caption{Pointwise variance in cartesian k-space per mask parameter over 10 randomly initialized runs on the fastMRI Knee dataset shown for various $\alpha$. Brighter color indicates a larger variance.}
    \label{fig:mask-std}
\end{figure}
\begin{figure}[b]
    \centering
    \addtolength{\tabcolsep}{-3pt}  
    \resizebox{\linewidth}{!}{
    \begin{tabular}{ccc}
        Original & Constr. mode of $\post$ & Random \post\ samples \\
         \includegraphics[width=0.4\linewidth]{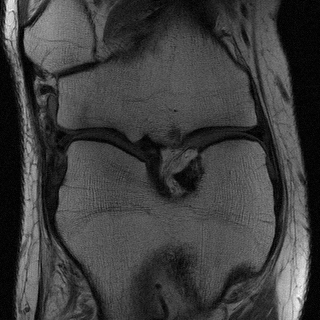} &  
         \includegraphics[width=0.4\linewidth]{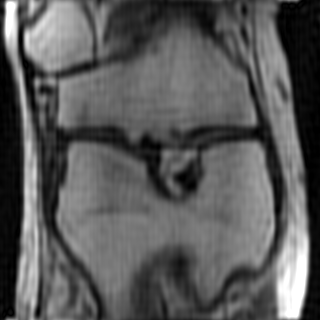} &  
         \includegraphics[width=0.4\linewidth]{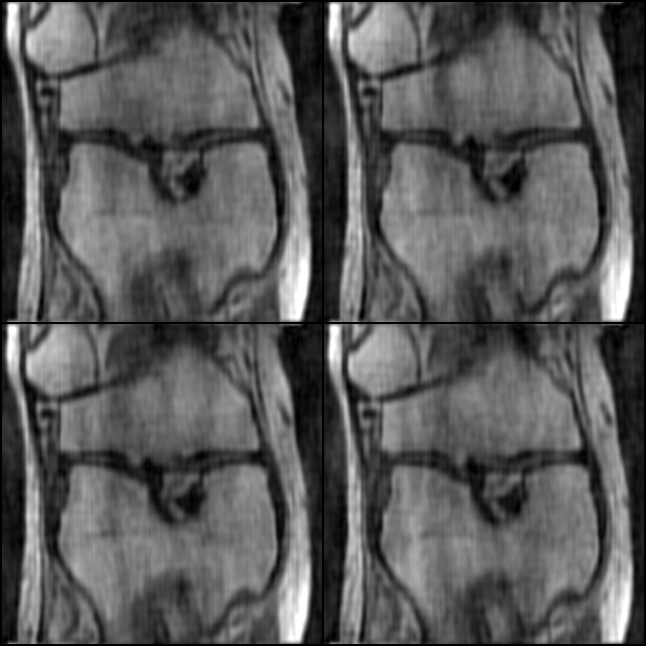}
    \end{tabular}
    }
    \addtolength{\tabcolsep}{3pt}  
    \caption{Comparison of an original scan (\textbf{left}) with reconstructions based on the constrained mode (\textbf{middle}) and random samples (\textbf{right}) of the learned mask distribution for fastMRI Knee.}
    \label{fig:mask-samples}
\end{figure}
\berm\ is a stochastic method, which involves randomness in its optimization process through the initialization of $\prior$ and Monte Carlo samples $\hat\x^{(l)}$ on each forward pass or batch collection of data samples.
Figure~\ref{fig:mask-std} displays the point-wise variance per mask parameter $\theta_i$ over multiple independent runs.
We observe the presence of a distinct central area that exhibits very small variance / high stability across all runs for most acceleration factors. 
Only in the case of an extreme acceleration factor (\texttt{x32}), more variance is found within this central region.
This suggests that particular segments within the k-space exhibit varying degrees of relevance, with the central region providing greater importance than the respective peripheral regions.
Consequently, an optimal sampling strategy would entail precisely targeting this central area.
If this region cannot be fully covered for larger acceleration factors ($\alpha = 32$), the main source of uncertainty is located within this important area with no distinct best solution.
\begin{figure}[t]
    \centering
    \addtolength{\tabcolsep}{-3pt}  
    \resizebox{\linewidth}{!}{
    \begin{tabular}{cccc}
        \texttt{x4} & \texttt{x8} & \texttt{x16} & \texttt{x32} \\
         \includegraphics[width=0.3\linewidth]{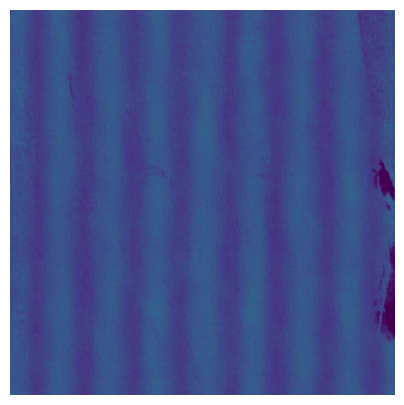} &  
         \includegraphics[width=0.3\linewidth]{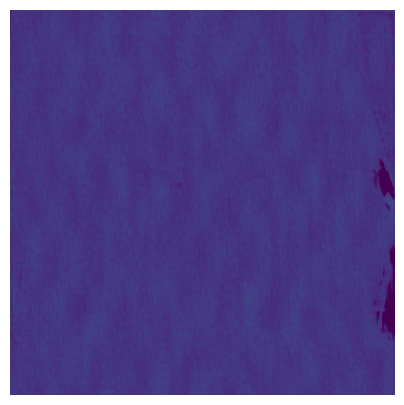} &  
         \includegraphics[width=0.3\linewidth]{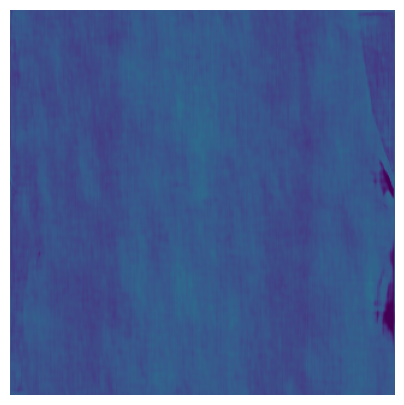} &  
         \includegraphics[width=0.3\linewidth]{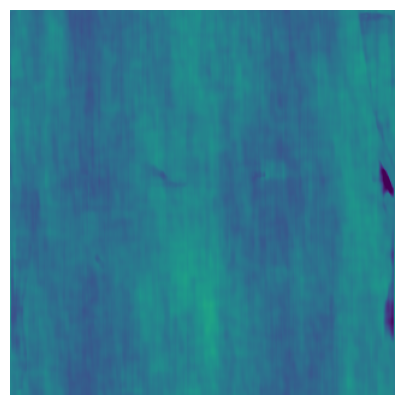}
    \end{tabular}
    }
    \addtolength{\tabcolsep}{3pt}  
    \caption{Pixel-variance on fastMRI Knee reconstructions based on random mask samples. Brighter color indicates larger variance.}
    \label{fig:mask-rec-std}
\end{figure}

Apart from the convergence of the mask distribution, sampling masks from $\post$ induces stochasticity in the inference process.
As shown in Figure~\ref{fig:mask-samples}, drawing random samples from $\post$ produces coherent images, which only differ slightly in their high-frequency noise patterns.
Interestingly, when investigating the variation of reconstructed pixels across multiple mask samples in Figure~\ref{fig:mask-rec-std}, a periodic pattern along the horizontal axis appears in the case of fastMRI Knee.
The wavelength of the observed pattern is getting longer with higher $\alpha$.
The stochasticity can be eliminated and reconstruction quality can be enhanced by using the suggested approach in Section~\ref{sec:postmode} and generating the mask based on the conditional posterior mode.

\subsection{Reconstruction Quality}

\begin{figure*}
    \centering
    \resizebox{\textwidth}{!}{
    \begin{tabular}{lc|c|c|c|c}
    
    & \textbf{Equispaced} & \textbf{2D Gaussian} & \textbf{IGS} & \textbf{\berm} & \textbf{Full} \\

    \tikz \node[rotate=90,]{\hspace{1mm} \textbf{\texttt{x16}}}; &

    \includegraphics[width=0.10\textwidth]{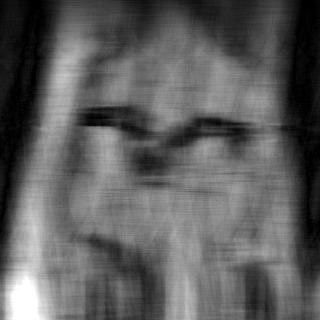}\hfill\hspace{0.5mm}\hfill \includegraphics[width=0.10\textwidth]{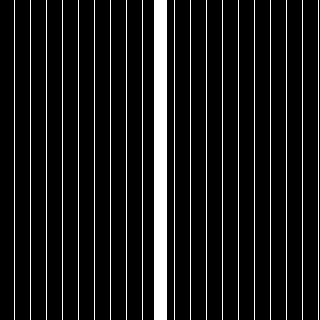} &
    
    \includegraphics[width=0.10\textwidth]{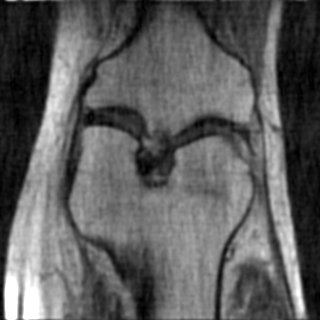}\hfill\hspace{0.5mm}\hfill
    \includegraphics[width=0.10\textwidth]{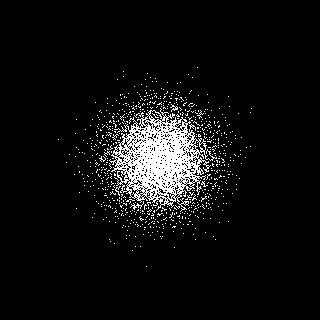} &

    \includegraphics[width=0.10\textwidth]{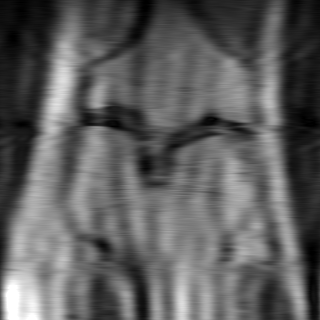}\hfill\hspace{0.5mm}\hfill 
    \includegraphics[width=0.10\textwidth]{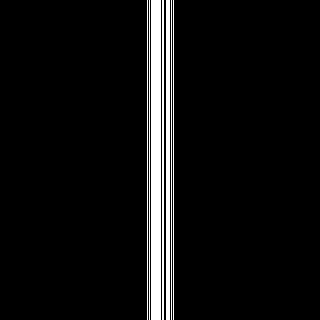} &

    \includegraphics[width=0.10\textwidth]{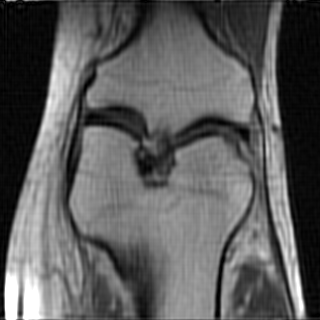}\hfill\hspace{0.5mm}\hfill 
    \includegraphics[width=0.10\textwidth]{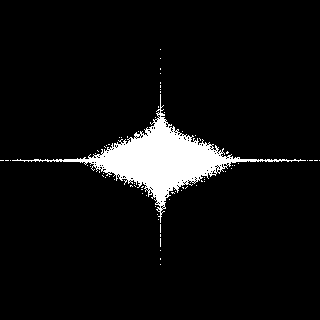} &

    \includegraphics[width=0.10\textwidth]{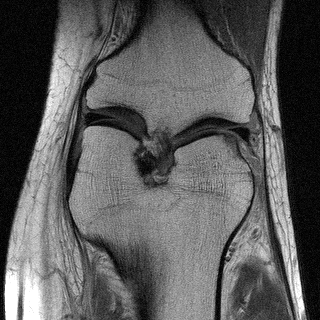}
    \\

    \tikz \node[rotate=90,]{\hspace{1mm} \textbf{\texttt{x32}}}; &

    \includegraphics[width=0.10\textwidth]{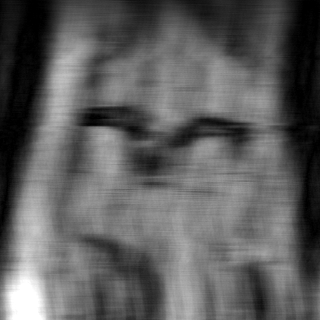}\hfill \includegraphics[width=0.10\textwidth]{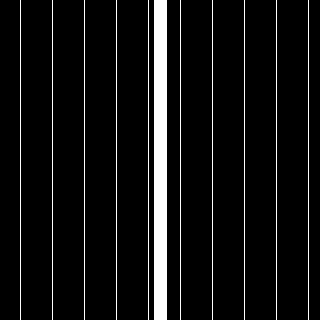} &
    
    \includegraphics[width=0.10\textwidth]{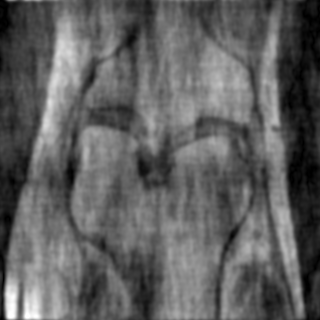}\hfill 
    \includegraphics[width=0.10\textwidth]{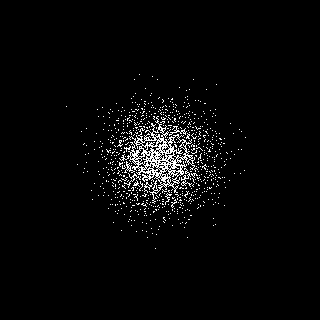} &

    \includegraphics[width=0.10\textwidth]{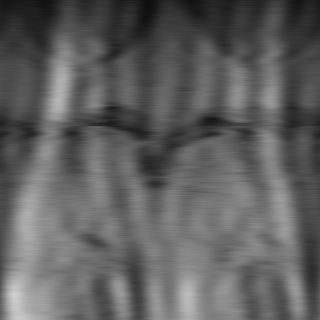}\hfill 
    \includegraphics[width=0.10\textwidth]{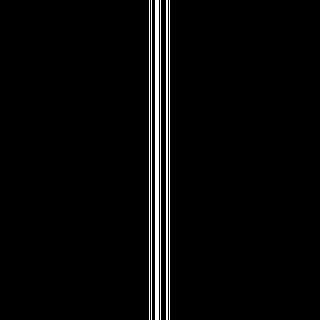} &

    \includegraphics[width=0.10\textwidth]{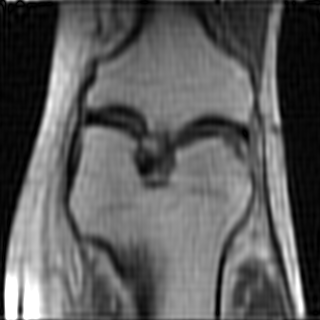}\hfill 
    \includegraphics[width=0.10\textwidth]{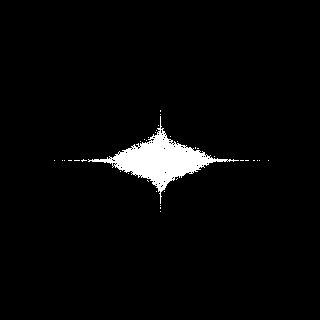} &

    \includegraphics[width=0.10\textwidth]{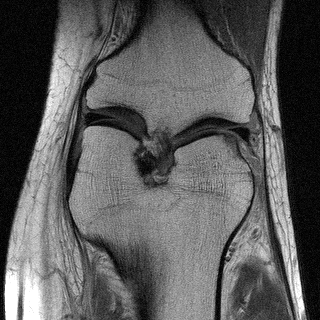}
    \\
    
    \end{tabular}
    }
    \caption{Reconstructions of a fastMRI Knee sample with $\alpha=16$ (\textbf{top}) and $\alpha=32$ (\textbf{bottom}).
    Each cell contains an evaluated method, where the reconstructed scan is displayed on the \textbf{left} and the applied mask on the \textbf{right} side.
    The original scan is shown in the \textbf{right} column.
    Equidistant spacing and IGS display strong infolding artifacts, which are less pronounced with Gaussian sampling. The reconstruction with the data-driven mask of \berm\ is much closer to the fully-sampled image and reduces image noise as well as artifacts very clearly.}
    \label{fig:ex-rec}
\end{figure*}
\begin{table*}
    \centering
    \resizebox{0.9\textwidth}{!}{ 
\begin{tabular}{clccc|ccc|ccc|ccc}
\toprule
        &  & \multicolumn{3}{c}{\textbf{\texttt{x4}}}                                                  & \multicolumn{3}{c}{\textbf{\texttt{x8}}}                                                  & \multicolumn{3}{c}{\textbf{\texttt{x16}}}                                                 & \multicolumn{3}{c}{\textbf{\texttt{x32}}} \\
\cmidrule(lr){3-5} \cmidrule(lr){6-8} \cmidrule(lr){9-11} \cmidrule(lr){12-14}

          & & \textit{PSNR} $\uparrow$ & \textit{SSIM} $\uparrow$ & \textit{NMSE} $\downarrow$ & \textit{PSNR} $\uparrow$ & \textit{SSIM} $\uparrow$ & \textit{NMSE} $\downarrow$ & \textit{PSNR} $\uparrow$ & \textit{SSIM} $\uparrow$ & \textit{NMSE} $\downarrow$ & \textit{PSNR} $\uparrow$ & \textit{SSIM} $\uparrow$ & \textit{NMSE} $\downarrow$ \\
\midrule
\parbox[t]{2mm}{\multirow{4}{*}{\rotatebox[origin=c]{90}{fastMRI}}} &
Equi.    & 24.481 & 0.601 & 0.066 & 23.803 & 0.524 & 0.077 & 23.365 & 0.474 & 0.085 & 23.158 & 0.448 & 0.090 \\
& Gauss.   & 29.570 & 0.664 & 0.027 & 28.118 & 0.560 & 0.035 & 23.342 & 0.440 & 0.112 & 17.373 & 0.299 & 0.396 \\
& IGS     & 28.553 & 0.640 & 0.031 & 26.428 & 0.532 & 0.047 & 24.376 & 0.458 & 0.070 & 22.313 & 0.409 & 0.108 \\
& \berm\   & \textbf{29.787} & \textbf{0.672} & \textbf{0.026} & \textbf{28.472} & \textbf{0.570} & \textbf{0.034} & \textbf{27.575} & \textbf{0.511} & \textbf{0.040} & \textbf{26.739} & \textbf{0.473} & \textbf{0.046}\\
\bottomrule

\parbox[t]{2mm}{\multirow{4}{*}{\rotatebox[origin=c]{90}{ACDC}}} &
Equi.    & 23.566 & 0.716 & 0.069 & 22.552 & 0.674 & 0.0857 & 21.785 & 0.653 & 0.103 & 21.772 & 0.650 & 0.104 \\
& Gauss.   & 39.169 & 0.959 & \textbf{0.001} & 24.681 & 0.705 & 0.073 & 18.038 & 0.512 & 0.265 & 15.786 & 0.428 & 0.422 \\
& IGS     & 33.856 & 0.925 & 0.007 & 27.804 & 0.813 & 0.026 & 23.804 & 0.712 & 0.065 & \textbf{20.705} & \textbf{0.621} & \textbf{0.132} \\
& \berm\   & \textbf{40.906} & \textbf{0.977} & \textbf{0.001} & \textbf{33.852} & \textbf{0.896} & \textbf{0.007} & \textbf{26.344} & \textbf{0.735} & \textbf{0.035} & 20.401 & 0.580 & 0.140\\
\bottomrule

\parbox[t]{2mm}{\multirow{4}{*}{\rotatebox[origin=c]{90}{BraTS}}} &
Equi.    & 26.732 & 0.716 & 0.033 & 25.599 & 0.687 & 0.043 & 25.660 & 0.693 & 0.042 & \textbf{25.149} & \textbf{0.692} & \textbf{0.048} \\
& Gauss.   & 38.372 & 0.779 & \textbf{0.002} & 22.452 & 0.301 & 0.112 & 17.543 & 0.225 & 0.307 & 14.357 & 0.152 & 0.574 \\
& IGS     & 35.550 & 0.911 & 0.004 & 30.607 & 0.821 & 0.014 & 26.995 & \textbf{0.739} & 0.032 & 24.054 & 0.679 & 0.061 \\
& \berm\   & \textbf{39.623} & \textbf{0.943} & \textbf{0.002} & \textbf{35.660} & \textbf{0.874} & \textbf{0.004} & \textbf{27.002} & 0.375 & \textbf{0.031} & 15.495 & 0.174 & 0.425\\
\bottomrule
\end{tabular}
}
\caption{Quality of reconstruction for the fastMRI Knee, ACDC, and BraTS.}
\label{tab:eval-rec}
\end{table*}
To assess the reconstruction quality of \berm\, we evaluate the peak signal-to-noise ratio (PSNR), structural similarity index measure (SSIM), and normalized mean squared error (NMSE) for acceleration factors $\alpha \in \{4,8,16,32\}$ (abbreviated as \texttt{x4} to \texttt{x32}) on the tests sets of fastMRI Knee, ACDC and BraTS.
Results (Table~\ref{tab:eval-rec}) show that the IGS method works well, the additional flexibility of \berm\ to operate in 2D, however, allows to obtain superior results.
For fastMRI Knee this advantage increases for higher $\alpha$.
As an instance, in the case of factor \texttt{x4}, the SSIM / NMSE of \berm\ is 5.00\% / 19.23\% better compared to IGS, with the improvement increasing to $15.64$\% / $134.78\%$ for factor \texttt{x32}.
Moreover, when compared to more conventional masking approaches, the gain in quality of \berm\ is even larger.
In addition, Figure~\ref{fig:ex-rec} shows that the \berm\ mask does not introduce noise nor artifacts, unlike the other evaluated methods.
However, even with a custom-tailored undersampling strategy, high-frequency details are omitted to a certain degree.

\begin{figure}[htbp]
    \centering
        \begin{tikzpicture}[
         image/.style = {text width=\linewidth, 
                         inner sep=0pt, outer sep=0pt},
        node distance = 1mm and 1mm
                                ] 
        \node [] (frame1)
            {
            \addtolength{\tabcolsep}{-1.5pt}  
            \begin{tabular}{ccc}
                 \includegraphics[trim={8mm 8mm 8mm 8mm},clip, width=0.2\linewidth]{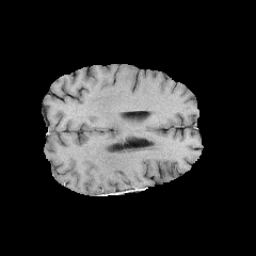} &  
                 \includegraphics[trim={8mm 8mm 8mm 8mm},clip, width=0.2\linewidth]{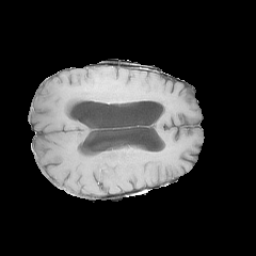} &  
                 \includegraphics[trim={8mm 8mm 8mm 8mm},clip, width=0.2\linewidth]{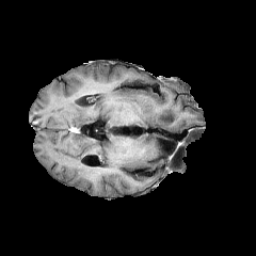} \\
                 \includegraphics[trim={8mm 8mm 8mm 8mm},clip, width=0.2\linewidth]{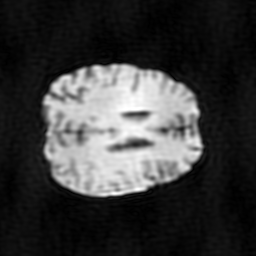} &  
                 \includegraphics[trim={8mm 8mm 8mm 8mm},clip, width=0.2\linewidth]{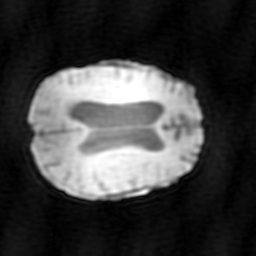} &  
                 \includegraphics[trim={8mm 8mm 8mm 8mm},clip, width=0.2\linewidth]{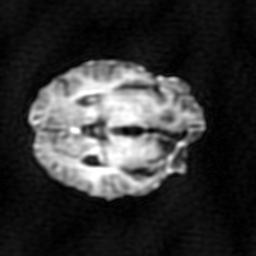}
            \end{tabular}
            \addtolength{\tabcolsep}{1.5pt}
            };
        
        \node [right=of frame1.west, xshift=-1mm, yshift=3.5mm, rotate=90] (frame2) {\scriptsize \textbf{Original}};
        \node [right=of frame1.west, xshift=-1mm, yshift=-18mm, rotate=90] (frame2) {\scriptsize \textbf{Reconstruction}};
    \end{tikzpicture}
 
    \caption{\berm\ reconstructions of T2-weighted BraTS samples with $\alpha = 16$ with high PSNR and low SSIM.}
    \label{fig:brats-psnr}
\end{figure}

On the other hand, when evaluating on the BraTS dataset, \berm\ is superior for acceleration factors \texttt{x4} and \texttt{x8}, but reveals performance limitations on higher $\alpha$ values.
Interestingly, for $\alpha = 16$ \berm\ exhibits the highest PSNR among the compared methods but has a rather low SSIM.
This can be reasoned by PSNR and SSIM capturing different aspects of image quality, where PSNR is primarily concerned with the amount of noise and SSIM focuses on structural information. 
Thus, this implies the presence of a low distortion level but differences in perceptual quality, as it is also visible in Figure~\ref{fig:brats-psnr}.

\subsection{Zero-shot Undersampled Segmentation}
\begin{figure*}
    \centering
    \resizebox{\textwidth}{!}{
    \begin{tabular}{lcccccc}
    
    & \textbf{Equispaced} & \textbf{2D Gaussian} & \textbf{IGS} & \textbf{\berm\ (1D)} & \textbf{\berm\ (2D)} & \textbf{Full} \\
    
    \tikz \node[rotate=90,]{\textbf{ACDC}}; &
    
    \includegraphics[width=0.10\textwidth]{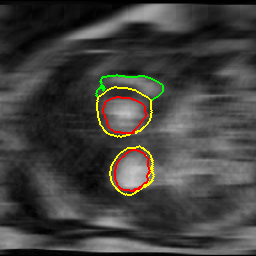}\hfill \includegraphics[width=0.10\textwidth]{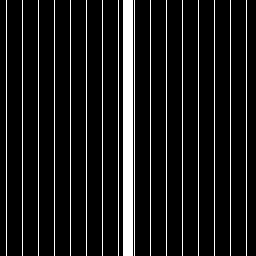} &
    
    \includegraphics[width=0.10\textwidth]{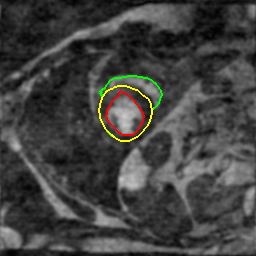}\hfill 
    \includegraphics[width=0.10\textwidth]{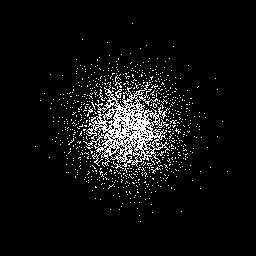} &

    \includegraphics[width=0.10\textwidth]{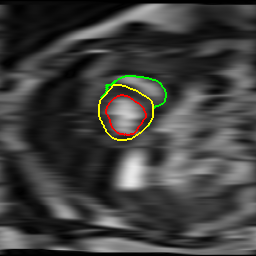}\hfill 
    \includegraphics[width=0.10\textwidth]{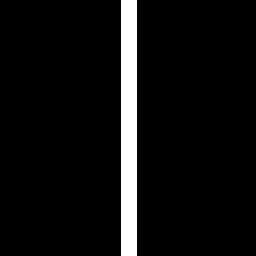} &

    \includegraphics[width=0.10\textwidth]{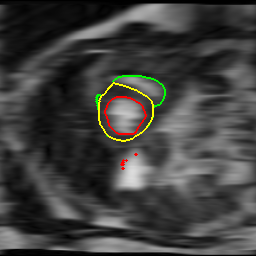}\hfill 
    \includegraphics[width=0.10\textwidth]{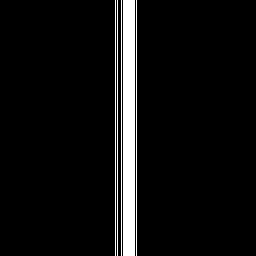} &

    \includegraphics[width=0.10\textwidth]{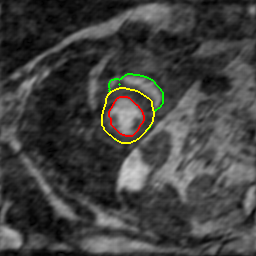}\hfill 
    \includegraphics[width=0.10\textwidth]{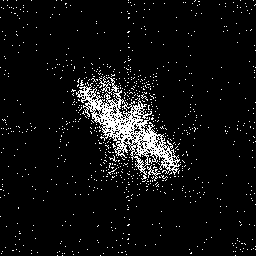} &

    \includegraphics[width=0.10\textwidth]{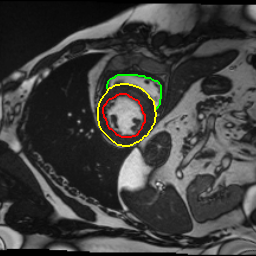}
    \\
    
    \midrule

    \tikz \node[rotate=90,]{\textbf{BraTS}}; &
    
    \includegraphics[trim={11mm 11mm 11mm 11mm},clip,width=0.10\textwidth]{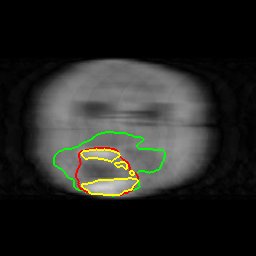}\hfill \includegraphics[width=0.10\textwidth]{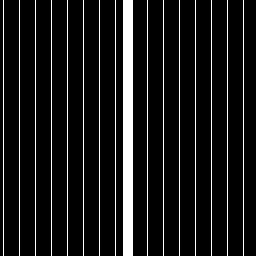} &
    
    \includegraphics[trim={11mm 11mm 11mm 11mm},clip,width=0.10\textwidth]{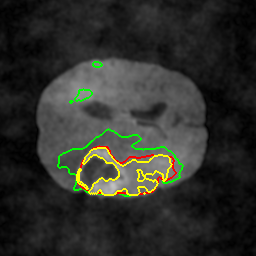}\hfill 
    \includegraphics[width=0.10\textwidth]{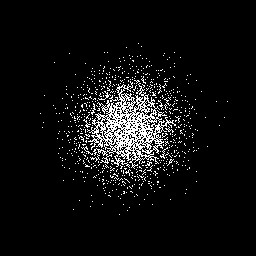} &

    \includegraphics[trim={11mm 11mm 11mm 11mm},clip,width=0.10\textwidth]{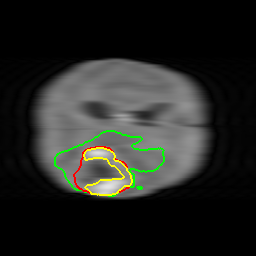}\hfill 
    \includegraphics[width=0.10\textwidth]{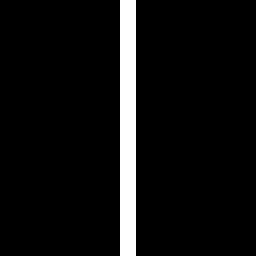} &

    \includegraphics[trim={11mm 11mm 11mm 11mm},clip,width=0.10\textwidth]{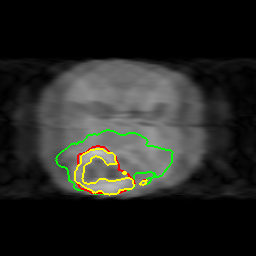}\hfill 
    \includegraphics[width=0.10\textwidth]{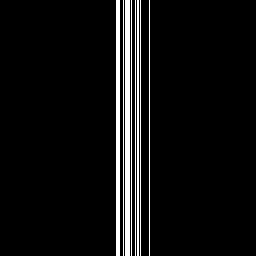} &

    \includegraphics[trim={11mm 11mm 11mm 11mm},clip,width=0.10\textwidth]{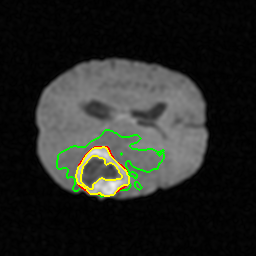}\hfill 
    \includegraphics[width=0.10\textwidth]{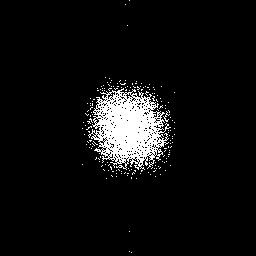} &

    \includegraphics[trim={11mm 11mm 11mm 11mm},clip,width=0.10\textwidth]{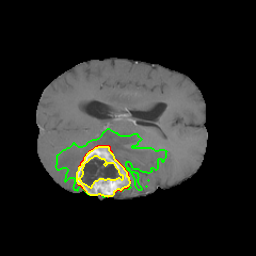}
    \\
    
    \end{tabular}
    }
    \vspace{-0.16cm}
    \caption{Segmentation with a pre-trained U-net using $\mathcal{L}_{\text{seg}}$ in \texttt{x16} accelerated \berm.
    The contours in ACDC correspond to the left (\colorbox{red!60}{\tiny \phantom{o}}) and right (\colorbox{green!60}{\tiny \phantom{o}}) ventricular cavity as well as the left ventricular myocardium (\colorbox{yellow!60}{\tiny \phantom{o}}). BraTS markings in the T1Gd sample imply whole (\colorbox{green!60}{\tiny \phantom{o}}), core (\colorbox{red!60}{\tiny \phantom{o}}) and enhancing (\colorbox{yellow!60}{\tiny \phantom{o}}) tumor.
    }
    \label{fig:ex-seg}
\end{figure*}

Visual reconstruction quality does not necessarily correlate with performance in downstream tasks such as segmentation \cite{razumov_optimal_2022}.
We now investigate the quality of \berm\ in a transfer learning scenario, where we apply a frozen segmentation network trained on fully-sampled MRIs and learn a mask to maximize its performance with undersampled MRIs.
For this, we choose a standard U-net \cite{ronneberger_u-net_2015} implementation and training routine with channel multipliers of (16, 32, 64, 128, 256).
We substitute $\mathcal{L}$ and use the trained segmentation network $net$ as a proxy paired with a combination of Dice and cross-entropy loss $\mathcal{L}_{\text{seg}}$, i.e. $\mathcal{L}(\hat{\x}, \x)$ becomes $\mathcal{L}_{\text{seg}}(net(\hat{\x}), \x_{\text{seg}})$ where $\x_{\text{seg}}$ is the segmentation label of \x.
Additionally, for the task of segmentation, we introduce a 1D variant of \berm\ to better understand its gain in performance compared to other methods.

The segmentation performance is evaluated in Table~\ref{tab:eval-seg} via Dice score and Intersection-over-Union (IuO).
Examples are visualized in Figure~\ref{fig:ex-seg}.
The fully-sampled MRI achieves a macro-averaged Dice score of 0.855 and an IoU of 0.763 for ACDC as well as 0.772 and 0.710 for BraTS, respectively.
While 1D \berm\ does not surpass the performance of IGS,
our contention is that 1D \berm\ would benefit from optimized parameters specifically tailored for 1D masking.
2D \berm\ achieves competitive segmentation performance for $\alpha = 8$ and is overall notably better for extreme acceleration factors such as $\alpha=32$ with a Dice score / IoU of $0.727$ / $0.606$ in ACDC and $0.706$ / $0.608$ in BraTS.
The Gaussian mask appears to be an efficient and straightforward method to achieve competitive performance for smaller $\alpha$.
Note that the \berm\ mask determined by $\mathcal{L}_{\text{seg}}$ differs substantially from the one obtained to maximize reconstruction quality (Figure~\ref{fig:ex-acc_fac}), emphasizing the fact that an optimal mask does not only vary with the dataset but also with the task.
\begin{table}[htbp]
    \centering
    \resizebox{\linewidth}{!}{ 
\newcolumntype{C}[1]{>{\centering\let\newline\\\arraybackslash\hspace{0pt}}m{#1}}
\begin{tabular}{cclcC{1cm}C{1cm}C{1cm}C{1cm}C{1cm}}
\toprule
  &   &       &     & Equi. & Gauss.  & IGS  & \berm\ (1D) & \berm\ (2D)\\
\midrule
\parbox[t]{2mm}{\multirow{6}{*}{\rotatebox[origin=c]{90}{ACDC}}} & \multirow{2}{*}{\texttt{x8}}  & \textit{Dice} & $\uparrow \,\,$ & 0.671 & \textbf{0.847} & 0.828 & 0.762 & 0.839\\
&                              & \textit{IoU} & $\uparrow \,\,$ & 0.546 & \textbf{0.752} & 0.726 & 0.650 & 0.742\\
\cmidrule(lr){2-9}
&\multirow{2}{*}{\texttt{x16}} & \textit{Dice} & $\uparrow \,\,$ & 0.645 & 0.644 & 0.745 & 0.717 & \textbf{0.789}\\
&                              & \textit{IoU} &  $\uparrow \,\,$  & 0.517 & 0.534 & 0.627 & 0.599 & \textbf{0.679} \\
\cmidrule(lr){2-9}
&\multirow{2}{*}{\texttt{x32}} & \textit{Dice} & $\uparrow \,\,$ & 0.644 & 0.399 & 0.592 & 0.587 & \textbf{0.727}\\
&                              & \textit{IoU} & $\uparrow \,\,$  & 0.517 & 0.301 & 0.466 & 0.460 & \textbf{0.606}\\
\toprule
\parbox[t]{2mm}{\multirow{6}{*}{\rotatebox[origin=c]{90}{BraTS}}} & \multirow{2}{*}{\texttt{x8}}  & \textit{Dice} & $\uparrow \,\,$  & 0.596 & 0.733 & 0.716 & 0.646 & \textbf{0.739} \\
&                              & \textit{IoU} & $\uparrow \,\,$ & 0.489 & 0.638 & 0.619 & 0.542 & \textbf{0.646} \\
\cmidrule(lr){2-9}
&\multirow{2}{*}{\texttt{x16}} & \textit{Dice} & $\uparrow \,\,$ & 0.589 & 0.597 & 0.651 & 0.537 & \textbf{0.735} \\
 &                             & \textit{IoU} &  $\uparrow \,\,$  & 0.481 & 0.494 & 0.546 & 0.426 & \textbf{0.640} \\
\cmidrule(lr){2-9}
&\multirow{2}{*}{\texttt{x32}} & \textit{Dice} & $\uparrow \,\,$ & 0.580 & 0.315 & 0.537 & 0.483 & \textbf{0.706} \\
&                              & \textit{IoU} & $\uparrow \,\,$ & 0.472 & 0.226 & 0.428 & 0.374 & \textbf{0.608} \\
\bottomrule
\end{tabular}
}
\caption{Segmentation metrics 
using a pre-trained U-net on ACDC and BraTS, showing that \berm\ (2D) excels especially for extreme acceleration factors.}
\label{tab:eval-seg}
\end{table}
%

\section{Conclusion} \label{sec:conclusion}

This paper proposes \berm\ as a general framework and building block for data-driven and probabilistic undersampling mask learning, with theoretically guaranteed acceleration factor enforcement.
Our evaluation protocol consisted in investigating image reconstruction and zero-shot segmentation of cardiac, brain, and knee MRIs.
\berm\ shows promising performance, with increased benefits compared to other methods, especially in the downstream task of segmentation, enabling extreme acceleration factors, which, can be traded into higher spatial resolution or coverage.
Further, \berm\ again demonstrated the benefit of not only having domain-specific but also task-specific undersampling masks, as the efficacy for a task does not necessarily overlap with visually correct reconstructions.

\paragraph{Clinical Practice.} 
MRI is an extremely flexible imaging modality, which involves optimization and multiple trade-offs, e.g., between total acquisition time (or temporal resolution in case of time-resolved imaging), spatial resolution and coverage, signal-to-noise ratio, and k-space data to be acquired.
Generally, any approach to reduce the amount of k-space data to be acquired while maintaining the quality of reconstructed images relaxes constraints on a multitude of protocol parameters and opens up possibilities for protocol optimization. 
Learning-based reconstruction and optimized k-space sampling for task-specific undersampling patterns as provided by \berm\ provides such an approach for better and faster MR data acquisition.
This is, e.g., particularly useful for high-speed interventions or pathology localization.
The minimalistic and model-free strategy of \berm\ allows us to derive these optimal masks within seconds from a single data sample.
As this also does not require extensive computational infrastructure, our approach further reduces barriers to actual deployment.

Currently, our evaluations have primarily focused on a 2D Cartesian pattern, forming scanner trajectories along the depth axis.
However, more scanner parameters and physical limitations need to be taken into account when considering deployment.
An extension of \berm\ to work with radial or spiral acquisition trajectories could open a new area of investigation.
Furthermore, the implications of deriving custom masks in a multi-coil setting need to be considered.
Additionally, custom masks for specific scanner parameters, acquisition protocols, or manufacturers could be developed.

\paragraph{Future Work.} 
Despite our take on not using a reconstruction network, \berm\ as an encapsulated layer with end-to-end differentiability can be integrated into any architecture of choice without major adaptions.
Further, potentially even higher performance could be achieved by finetuning the network of the respective downstream task, e.g., segmentation, during the \berm\ process.
Another possible direction to improve our approach is to integrate prior domain knowledge into $\prior$.
While our factorized prior distribution assumption proved to work well in most experiments, it could be beneficial to incorporate further prior knowledge if data is scarce, i.e., for very large acceleration factors.

\section*{Acknowledgments}

This work has been partially funded by the Deutsche Forschungsgemeinschaft (DFG, German Research Foundation) as part of BERD@NFDI - grant number 460037581.
The authors gratefully acknowledge LMU Klinikum for providing computing resources on their Clinical Open Research Engine (CORE).

{\small
\bibliographystyle{ieee_fullname}
\bibliography{bernoulli-mask}

\begin{thebibliography}{10}\itemsep=-1pt

\bibitem{bahadir_learning-based_2019}
Cagla~Deniz Bahadir, Adrian~V Dalca, and Mert~R Sabuncu.
\newblock Learning-based optimization of the under-sampling pattern in {MRI}.
\newblock In {\em Information Processing in Medical Imaging: 26th International
  Conference, IPMI}, pages 780--792, 2019.

\bibitem{bakas_advancing_2017}
Spyridon Bakas, Hamed Akbari, Aristeidis Sotiras, Michel Bilello, Martin
  Rozycki, et~al.
\newblock Advancing the cancer genome atlas glioma mri collections with expert
  segmentation labels and radiomic features.
\newblock {\em Scientific data}, 4(1):1--13, 2017.

\bibitem{bakas_identifying_2019}
Spyridon Bakas, Mauricio Reyes, Andras Jakab, Stefan Bauer, Markus Rempfler,
  et~al.
\newblock Identifying the best machine learning algorithms for brain tumor
  segmentation, progression assessment, and overall survival prediction in the
  brats challenge.
\newblock {\em arXiv:1811.02629 [cs, stat]}, 2018.

\bibitem{belov_towards_2021}
Aleksandr Belov, Jo{\"e}l Stadelmann, Sergey Kastryulin, and Dmitry~V Dylov.
\newblock Towards ultrafast mri via extreme k-space undersampling and
  superresolution.
\newblock In {\em Medical Image Computing and Computer Assisted Intervention,
  MICCAI}, pages 254--264, 2021.

\bibitem{bernard_deep_2018}
Olivier Bernard, Alain Lalande, Clement Zotti, Frederick Cervenansky, Xin Yang,
  Pheng-Ann Heng, Irem Cetin, Karim Lekadir, Oscar Camara, Miguel
  Angel~Gonzalez Ballester, et~al.
\newblock Deep learning techniques for automatic mri cardiac multi-structures
  segmentation and diagnosis: is the problem solved?
\newblock {\em IEEE Transactions on Medical Imaging}, 37(11):2514--2525, 2018.

\bibitem{bian_learnable_2022}
Wanyu Bian, Qingchao Zhang, Xiaojing Ye, and Yunmei Chen.
\newblock A learnable variational model for joint multimodal {MRI}
  reconstruction and synthesis.
\newblock In {\em Medical Image Computing and Computer Assisted Intervention,
  MICCAI}, pages 354--364, 2022.

\bibitem{campbell2017real}
Adrienne~E Campbell-Washburn, Mohammad~A Tavallaei, Mihaela Pop, Elena~K Grant,
  Henry Chubb, Kawal Rhode, and Graham~A Wright.
\newblock Real-time mri guidance of cardiac interventions.
\newblock {\em Journal of Magnetic Resonance Imaging}, 46(4):935--950, 2017.

\bibitem{chung_score-based_2022}
Hyungjin Chung and Jong~Chul Ye.
\newblock Score-based diffusion models for accelerated mri.
\newblock {\em Medical Image Analysis}, 80:102479, 2022.

\bibitem{gungor_adaptive_2022}
Salman~UH Dar, {\c{S}}aban {\"O}zt{\"u}rk, Yilmaz Korkmaz, Gokberk Elmas,
  Muzaffer {\"O}zbey, et~al.
\newblock Adaptive diffusion priors for accelerated mri reconstruction.
\newblock {\em arXiv:2207.05876 [cs, eess]}, 2022.

\bibitem{deshmane2012parallel}
Anagha Deshmane, Vikas Gulani, Mark~A Griswold, and Nicole Seiberlich.
\newblock Parallel mr imaging.
\newblock {\em Journal of Magnetic Resonance Imaging}, 36(1):55--72, 2012.

\bibitem{ding_mri_2022}
Qiaoqiao Ding and Xiaoqun Zhang.
\newblock Mri reconstruction by completing under-sampled k-space data with
  learnable fourier interpolation.
\newblock In {\em Medical Image Computing and Computer Assisted Intervention,
  MICCAI}, pages 676--685, 2022.

\bibitem{dong_invertible_2022}
Siyuan Dong, Eric~Z Chen, Lin Zhao, Xiao Chen, Yikang Liu, Terrence Chen, and
  Shanhui Sun.
\newblock Invertible sharpening network for mri reconstruction enhancement.
\newblock In {\em Medical Image Computing and Computer Assisted Intervention,
  MICCAI}, pages 582--592, 2022.

\bibitem{gao_projection-based_2022}
Chang Gao, Shu-Fu Shih, J~Paul Finn, and Xiaodong Zhong.
\newblock A projection-based k-space transformer network for undersampled
  radial mri reconstruction with limited training subjects.
\newblock In {\em Medical Image Computing and Computer Assisted Intervention,
  MICCAI}, pages 726--736, 2022.

\bibitem{heidemann2003brief}
Robin~M Heidemann, {\"O}zkan {\"O}zsarlak, Paul~M Parizel, Johan Michiels,
  Berthold Kiefer, Vladimir Jellus, Mathias M{\"u}ller, Felix Breuer, Martin
  Blaimer, Mark~A Griswold, et~al.
\newblock A brief review of parallel magnetic resonance imaging.
\newblock {\em European radiology}, 13:2323--2337, 2003.

\bibitem{jang_categorical_2017}
Eric Jang, Shixiang Gu, and Ben Poole.
\newblock Categorical reparameterization with gumbel-softmax.
\newblock In {\em 5th International Conference on Learning Representations,
  {ICLR}}, 2017.

\bibitem{kaplan2002real}
Irving Kaplan, Nicklas~E Oldenburg, Paul Meskell, Michael Blake, Paul Church,
  and Edward~J Holupka.
\newblock Real time mri-ultrasound image guided stereotactic prostate biopsy.
\newblock {\em Magnetic resonance imaging}, 20(3):295--299, 2002.

\bibitem{kingma_adam_2017}
Diederik~P. Kingma and Jimmy Ba.
\newblock Adam: {A} method for stochastic optimization.
\newblock In {\em 3rd International Conference on Learning Representations,
  {ICLR}}, 2015.

\bibitem{kingma_auto-encoding_2022}
Diederik~P. Kingma and Max Welling.
\newblock Auto-encoding variational bayes.
\newblock In {\em 2nd International Conference on Learning Representations,
  {ICLR}}, 2014.

\bibitem{lazarus2019sparkling}
Carole Lazarus, Pierre Weiss, Nicolas Chauffert, Franck Mauconduit, Loubna
  El~Gueddari, Christophe Destrieux, Ilyess Zemmoura, Alexandre Vignaud, and
  Philippe Ciuciu.
\newblock Sparkling: variable-density k-space filling curves for accelerated
  t2*-weighted mri.
\newblock {\em Magnetic resonance in medicine}, 81(6):3643--3661, 2019.

\bibitem{liu_undersampled_2022}
Xinwen Liu, Jing Wang, Cheng Peng, Shekhar~S Chandra, Feng Liu, and S~Kevin
  Zhou.
\newblock Undersampled mri reconstruction with side information-guided
  normalisation.
\newblock In {\em Medical Image Computing and Computer Assisted Intervention,
  MICCAI}, pages 323--333, 2022.

\bibitem{lustig_sparse_2007}
Michael Lustig, David Donoho, and John~M Pauly.
\newblock Sparse {MRI}: The application of compressed sensing for rapid mr
  imaging.
\newblock {\em Magnetic Resonance in Medicine}, 58(6):1182--1195, 2007.

\bibitem{menze_multimodal_2015}
Bjoern~H Menze, Andras Jakab, Stefan Bauer, Jayashree Kalpathy-Cramer, Keyvan
  Farahani, Justin Kirby, Yuliya Burren, Nicole Porz, Johannes Slotboom, Roland
  Wiest, et~al.
\newblock The multimodal brain tumor image segmentation benchmark ({BRATS}).
\newblock {\em IEEE Transactions on Medical Imaging}, 34(10):1993--2024, 2014.

\bibitem{nayak2022real}
Krishna~S Nayak, Yongwan Lim, Adrienne~E Campbell-Washburn, and Jennifer
  Steeden.
\newblock Real-time magnetic resonance imaging.
\newblock {\em Journal of Magnetic Resonance Imaging}, 55(1):81--99, 2022.

\bibitem{oztek2020practical}
Murat~Alp Oztek, Christina~L Brunnquell, Michael~N Hoff, Daniel~J Boulter,
  Mahmud Mossa-Basha, Luke~H Beauchamp, David~L Haynor, and Xuan~V Nguyen.
\newblock Practical considerations for radiologists in implementing a
  patient-friendly mri experience.
\newblock {\em Topics in Magnetic Resonance Imaging}, 29(4):181--186, 2020.

\bibitem{paszke_pytorch_2019}
Adam Paszke, Sam Gross, Francisco Massa, Adam Lerer, James Bradbury, Gregory
  Chanan, Trevor Killeen, Zeming Lin, Natalia Gimelshein, Luca Antiga, et~al.
\newblock Pytorch: An imperative style, high-performance deep learning library.
\newblock {\em Advances in neural information processing systems}, 32, 2019.

\bibitem{peng_towards_2022}
Cheng Peng, Pengfei Guo, S~Kevin Zhou, Vishal~M Patel, and Rama Chellappa.
\newblock Towards performant and reliable undersampled mr reconstruction via
  diffusion model sampling.
\newblock In {\em Medical Image Computing and Computer Assisted Intervention,
  MICCAI}, pages 623--633, 2022.

\bibitem{razumov_optimal_2022}
Artem Razumov, Oleg~Y Rogov, and Dmitry~V Dylov.
\newblock Optimal {MRI} undersampling patterns for ultimate benefit of medical
  vision tasks.
\newblock {\em arXiv:2108.04914 [cs, eess]}, 2021.

\bibitem{razumov_optimal_2022-1}
Artem Razumov, Oleg~Y Rogov, and Dmitry~V Dylov.
\newblock Optimal mri undersampling patterns for pathology localization.
\newblock In {\em Medical Image Computing and Computer Assisted Intervention,
  MICCAI}, pages 768--779, 2022.

\bibitem{ronneberger_u-net_2015}
Olaf Ronneberger, Philipp Fischer, and Thomas Brox.
\newblock U-net: Convolutional networks for biomedical image segmentation.
\newblock In {\em Medical Image Computing and Computer-Assisted Intervention,
  MICCAI}, pages 234--241, 2015.

\bibitem{uecker2010real}
Martin Uecker, Shuo Zhang, Dirk Voit, Alexander Karaus, Klaus-Dietmar Merboldt,
  and Jens Frahm.
\newblock Real-time mri at a resolution of 20 ms.
\newblock {\em NMR in Biomedicine}, 23(8):986--994, 2010.

\bibitem{wang_b-spline_2022}
Guanhua Wang, Tianrui Luo, Jon-Fredrik Nielsen, Douglas~C Noll, and Jeffrey~A
  Fessler.
\newblock B-spline parameterized joint optimization of reconstruction and
  k-space trajectories (bjork) for accelerated 2d mri.
\newblock {\em IEEE Transactions on Medical Imaging}, 41(9):2318--2330, 2022.

\bibitem{weiss2019pilot}
Tomer Weiss, Ortal Senouf, Sanketh Vedula, Oleg Michailovich, Michael
  Zibulevsky, and Alex Bronstein.
\newblock Pilot: Physics-informed learned optimized trajectories for
  accelerated mri.
\newblock {\em arXiv:1909.05773}, 2019.

\bibitem{weiss_joint_2020}
Tomer Weiss, Sanketh Vedula, Ortal Senouf, Oleg Michailovich, Michael
  Zibulevsky, and Alex Bronstein.
\newblock Joint learning of cartesian under sampling andre construction for
  accelerated mri.
\newblock In {\em IEEE International Conference on Acoustics, Speech and Signal
  Processing (ICASSP)}, pages 8653--8657. IEEE, 2020.

\bibitem{xie_puert_2022}
Jingfen Xie, Jian Zhang, Yongbing Zhang, and Xiangyang Ji.
\newblock Puert: Probabilistic under-sampling and explicable reconstruction
  network for cs-mri.
\newblock {\em IEEE Journal of Selected Topics in Signal Processing},
  16(4):737--749, 2022.

\bibitem{xue20222d}
Shengke Xue, Zhaowei Cheng, Guangxu Han, Chaoliang Sun, Ke Fang, et~al.
\newblock 2d probabilistic undersampling pattern optimization for mr image
  reconstruction.
\newblock {\em Medical Image Analysis}, 77:102346, 2022.

\bibitem{zbontar_fastmri_2019}
Jure Zbontar, Florian Knoll, Anuroop Sriram, Tullie Murrell, Zhengnan Huang,
  Matthew~J Muckley, Aaron Defazio, Ruben Stern, Patricia Johnson, Mary Bruno,
  et~al.
\newblock fastmri: An open dataset and benchmarks for accelerated mri.
\newblock {\em arXiv:1811.08839 [physics, stat]}, 2019.

\bibitem{zhao_fastmri_2021}
Ruiyang Zhao, Burhaneddin Yaman, Yuxin Zhang, Russell Stewart, Austin Dixon,
  Florian Knoll, Zhengnan Huang, Yvonne~W Lui, Michael~S Hansen, and Matthew~P
  Lungren.
\newblock fastmri+: Clinical pathology annotations for knee and brain fully
  sampled multi-coil mri data.
\newblock {\em arXiv:2109.03812 [physics]}, 2021.

\bibitem{zhou_effective_2021}
Xiao Zhou, Weizhong Zhang, Hang Xu, and Tong Zhang.
\newblock Effective sparsification of neural networks with global sparsity
  constraint.
\newblock In {\em Proceedings of the IEEE/CVF Conference on Computer Vision and
  Pattern Recognition}, pages 3599--3608, 2021.

\bibitem{zhu_prune_2017}
Michael Zhu and Suyog Gupta.
\newblock To prune, or not to prune: Exploring the efficacy of pruning for
  model compression.
\newblock In {\em 6th International Conference on Learning Representations,
  {ICLR}, Workshop Track Proceedings}, 2018.

\end{thebibliography}
}

\appendix
\clearpage
\section{Supplementary Material}

\subsection{$S$ annealing schedule} \label{app:anneal}

Acceleration factor $\alpha$, annealing start iteration $i_{min}$, annealing end iteration $i_{max}$, current iteration $i_{cur}$, $d_{target} = \frac{1}{\alpha}$.
\begin{align*} 
d_{cur} &= d_{target} + (1 - d_{target}) \left( 1 - \frac{i_{cur} - i_{min}}{i_{max} - i_{min}} \right) \\
S_{cur} &= d_{cur} \cdot D
\end{align*}

\subsection{Proof of Constraint Projection} \label{app:proj}

Proof for Eq.~\ref{eq:projdesc} and Eq.~\ref{eq:projeq} is taken and adapted from \cite{zhou_effective_2021}.
Transforming updated parameters $\tilde{\s} \in \mathbb{R}^D$ into \s, which fulfills the sparsification constraint can be described as a least-squares convex problem:
\begin{equation}
       \argmin_{\s} \frac{1}{2} \lvert\lvert \tilde{\s} - \s \rvert\rvert^2\:\:\:\:\:\:
   s.t. \:  \sum_{i=1}^D \theta_i = \bm 1^\top \s \leq S\: \text{and} \: 0 \leq \theta_i \leq 1 \eqendp
\end{equation}
This can be solved by the Lagrangian multiplier method:
\begin{align}
    \mathfrak{L}(\s, \lambda) &= \frac{1}{2} \lvert\lvert \s - \tilde{\s} \rvert\rvert^2 + \lambda(\bm 1^\top \s - S) \\
    &= \frac{1}{2} \lvert\lvert \s - (\tilde{\s} - \lambda \bm 1) \rvert\rvert^2 + \lambda(\bm 1^\top \tilde{\s} - S) - \frac{n}{2} \lambda^2 \eqendc
\end{align}
where $\lambda \geq 0$ and $0 \leq \theta_i \leq 1$. Minimizing w.r.t. \s\ results in

\begin{equation}
    \bar \s = \bm 1_{\tilde{s} - \lambda \bm 1 \geq 1} + (\tilde{s} - \lambda \bm 1)_{1 > \tilde{s} - \lambda \bm 1 > 0} \eqendp
\end{equation}

Thus, for $\lambda \geq 0$ 
\begin{equation}
    \begin{split}
        g(\lambda) &= \mathfrak{L}(\bar \s, \lambda) \\
    &= \frac{1}{2} \lvert\lvert [\tilde{s} - \lambda \bm 1]_{-} + [\tilde{\s} - (\lambda + 1) \bm 1]_{+} \rvert\rvert^2  \\ & + \lambda(\bm 1^\top \tilde{\s} - \s) - \frac{D}{2} \lambda^2 \\
    &= \frac{1}{2} \lvert\lvert [\tilde{s} - \lambda \bm 1]_{-} \rvert\rvert + \frac{1}{2} \lvert\lvert [\tilde{\s} - (\lambda + 1) \bm 1]_{+} \rvert\rvert^2 \\ & + \lambda(\bm 1^\top \tilde{\s} - \s) - \frac{D}{2} \lambda^2    
    \end{split}
\end{equation}
and
\begin{equation}
    \begin{split}    g'(\lambda) &= \bm 1^\top [\lambda \bm 1 - \tilde{\s}]_{+} + \bm 1^\top [(\lambda + 1) \bm 1 - \s]_{-} \\ & + (\bm 1^\top \tilde{\s} - \s) - D \lambda \\
    &= \bm 1^\top \min(\bm 1, \max(\bm 0, \tilde{\s} - \lambda \bm 1)) - S \\
    &= [\textstyle \sum_{i=1}^D \min (1, \max(0, \tilde{\theta_i} - \lambda))] - S \eqendp
    \end{split}
\end{equation}

With $g'(\lambda)$ being a monotone function, $\lambda_1^*$ a solution for $g'(\lambda) = 0$ can be obtained by e.g. a convex solver or a bisection method.
The maximum of $g(\lambda)$ with $\lambda \geq 0$ is at $\lambda_2^* = \max(0, \lambda_1^*)$.
Eventually,

\begin{align}
        \s^* &= \bm 1_{\tilde{s} - \lambda_2^* \bm 1 \geq 1} + (\tilde{s} - \lambda_2^* \bm 1)_{1 > \tilde{s} - \lambda_2^* \bm 1 > 0} \\
        &= \min ( \bm 1, \max ( \bm 0, \tilde{\s} -  \lambda_2^* \bm 1) \\ 
        &= \min ( \bm 1, \max ( \bm 0, \tilde{\s} -  \max(0, \lambda_1^*) \bm 1) \eqendp
\end{align}

\end{document}